\documentclass[graybox,natbib,nosecnum]{svmult}
\bibpunct{(}{)}{;}{a}{}{,} 

\pdfoutput=1   

\usepackage{mathptmx}       
\usepackage{helvet}         
\usepackage{courier}        
\usepackage{type1cm}        

\usepackage{makeidx}         
\usepackage{graphicx}        
\usepackage{multicol}        
\usepackage[bottom]{footmisc}
\usepackage[normalem]{ulem}	
\usepackage{hyperref}  

\usepackage{soul}   

\newcommand*\aap{A\&A}

\newcommand*\aj{AJ}

\newcommand*\apj{ApJ}

\newcommand*\mnras{MNRAS}

\newcommand*\pasp{PASP}

\newcommand*\procspie{Proc SPIE}

\usepackage{xspace}

\newcommand{\simi}{\ensuremath{\sim}}

\newcommand{\degreesq}{\ensuremath{^{\circ2}}\xspace}

\newcommand{\ie}{\textit{i.e.}\xspace}
\newcommand{\kepler}{\textit{Kepler}\xspace}
\newcommand{\corot}{\textit{CoRoT}\xspace}
\newcommand{\tess}{\textit{TESS}\xspace}

\makeindex             

\begin{document}

\title*{Small Telescope Exoplanet Transit Surveys: XO}
\author{Nicolas Crouzet}
\institute{Nicolas Crouzet \at Instituto de Astrof\'isica de Canarias, C. V\'ia L\'actea S/N, E-38205 La Laguna, Tenerife, Spain, \email{ncrouzet@iac.es}}
%
%
\maketitle

\abstract{}

The XO project aims at detecting transiting exoplanets around bright stars from the ground using small telescopes. The original configuration of XO \citep{McCullough2005} has been changed and extended as described here. The instrumental setup consists of three identical units located at different sites, each composed of two lenses equipped with CCD cameras mounted on the same mount. We observed two strips of the sky covering an area of 520\degreesq for twice nine months. We build lightcurves for \simi20,000 stars up to magnitude $R \approx 12.5$ using a custom-made photometric data reduction pipeline. The photometric precision is around 1-2\% for most stars, and the large quantity of data allows us to reach a millimagnitude precision when folding the lightcurves on timescales that are relevant to exoplanetary transits. We search for periodic signals and identify several hundreds of variable stars and a few tens of transiting planet candidates. Follow-up observations are underway to confirm or reject these candidates. We found two close-in gas giant planets so far, in line with the expected yield.

\section{Introduction}

Much of our knowledge of gas giant exoplanets comes from those transiting in front of bright stars on close-in orbits. These planets are easier to detect and are more favorable to follow-up observations such as radial velocity measurements. Their atmospheric characterization through transit spectroscopy is also facilitated by the large amount of flux received from bright stars. The vast majority of these planets have been discovered by dedicated ground-based photometric surveys, the most prolific being WASP \citep{Pollacco2006, Collier2007} and HAT \citep{Bakos2002, Bakos2004}. These two projects detected \simi80 hot Jupiters around stars of magnitude $V \leq 11$. The \corot and \kepler missions provided little addition to this sample because they target fainter stars and much smaller sky areas.
The XO project \citep{McCullough2005} aims at detecting transiting exoplanets around bright stars from the ground using small telescopes. The project started in 2003 and discovered five close-in gas giant planets, XO-1b to XO-5b \citep{McCullough2006, Burke2007, JohnsKrull2008, McCullough2008, Burke2008}. A second version of XO was deployed in 2011 and 2012 and operated nominally from 2012 to 2014. This yielded the discovery of the hot Jupiter XO-6b orbiting a bright, hot, and fast rotating star \citep{Crouzet2017}, and of other planet candidates.
In this chapter, we provide an overview of this second version of XO. First, we present the goals of the project, the instrumental setup, and the observation strategy. Then, we detail the various steps of the data reduction pipeline, and we review the instrumental performances. Finally, we present the search for periodic signals in the lightcurves, the identification and follow-up observations of transit candidates, and we compare the detection yield to the number of detected planets.

\section{Goals of the project}
\label{sec: Goals}

The XO project aims at detecting transiting exoplanets around bright stars ($V < 11$) including some with long orbital periods ($P > 10$ days). Bright host stars ensure that follow-up observations can be achieved to confirm the planet candidates, and make their atmospheric characterization possible. Besides, whereas close-in transiting giant planets have been discovered and studied by dozens, only a few transiting ones with long orbital periods and bright host stars are known. The long period severely decreases the transit probability making such detections challenging. As a result, constraints on the physical properties of long period giant exoplanets have been obtained only recently from a statistical study of \kepler objects, and their atmospheric properties are largely unknown \citep[see][and the chapter entitled \underline{Hot Jupiter Populations from Transit and RV Surveys} of this book]{Santerne2016}.

\section{Instrumental setup}
\label{sec: Instrumental setup}

\begin{figure}
\includegraphics[width=5cm]{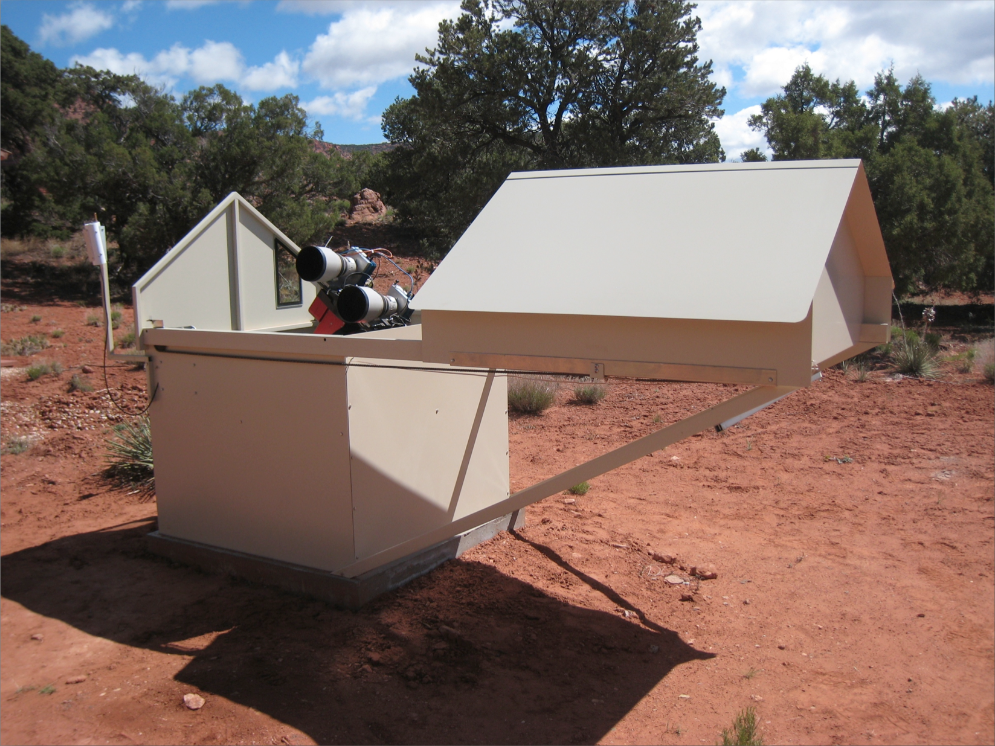}
\hfill
\includegraphics[width=5.7cm]{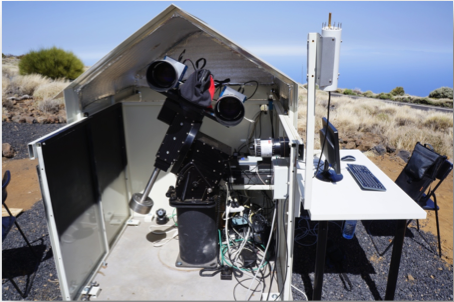}
\caption{XO units at Vermillion Cliffs Observatory, Kanab, Utah (left) and at the Observatorio del Teide, Canary Islands (right). A third unit is installed at the Observatori Astron\`omic del Montsec near \`Ager, mainland Spain.}
\label{fig: XO units}
\end{figure}

XO is composed of three identical units installed at Vermillion Cliffs Observatory, Kanab, Utah, at Observatorio del Teide, Tenerife, Canary Islands, and at Observatori Astron\`omic del Montsec, near \`Ager, Spain. Each unit is composed of two 10 cm diameter and 200 mm focal length Canon telephoto lenses equipped with Apogee E6 1024$\times$1024 pixels CCD cameras, mounted on a German-Equatorial Paramount ME mount, and protected by a shelter with a computer-controlled roof (Figure \ref{fig: XO units}). 
The basic hardware is described in \citet{McCullough2005} but was changed a bit in the transition from the first version of XO at a single observatory at Haleakala, Hawaii, to the three observatories in the small roll-off roof enclosures. The main change was from back-illuminated SITe CCDs with parallel port interface to front-side illuminated Kodak CCDs with Ethernet interface, as a cost compromise (the manufacturer SITe stopped making the inexpensive back-side illuminated CCDs).
All six lenses and cameras operate in a network configuration: they point toward the same fields of view over the night at the three locations. The CCDs are read in time delayed integration (TDI): pixels are read continuously while stars move along columns on the detector. The scan and read rates are adjusted to maintain round PSFs (Point Spread Functions). This setup results in strips of $43^{\circ}.2\times7^{\circ}.2$ on the sky instead of square images. This technique maximizes the number of observed bright stars by enlarging the effective field of view. The exposure time to acquire a full strip is 5.3 minutes, which corresponds to 53 seconds on each $7^{\circ}.2\times7^{\circ}.2$ square subimage. The nominal PSF FWHM (Full Width Half Maximum) is 1.2 pixels and varies in practice between 1 and 2.5 pixels (see Section \textit{\nameref{sec: FWHM variations}}). The units can be controlled remotely and operate robotically. Data from a weather sensor are recorded and analyzed in real time; they are used to send instructions such as opening or closing the roof, running the observations, etc. A webcam sensitive in the optical and infrared is mounted inside each enclosure and can be accessed from a web interface. A computer running on Linux is installed inside each unit to control the operations and store the data. An IDL program commands the operations of the unit and runs the observations. This program as well as other daily tasks are launched using crontab.

\newpage

\section{Observations}
\label{sec: Observations}

This section presents the observation strategy that we adopted for XO and gives an overview of the operations.

\subsection{Observation strategy}
\label{sec: Observation strategy}

\begin{figure}
\includegraphics[width=\columnwidth]{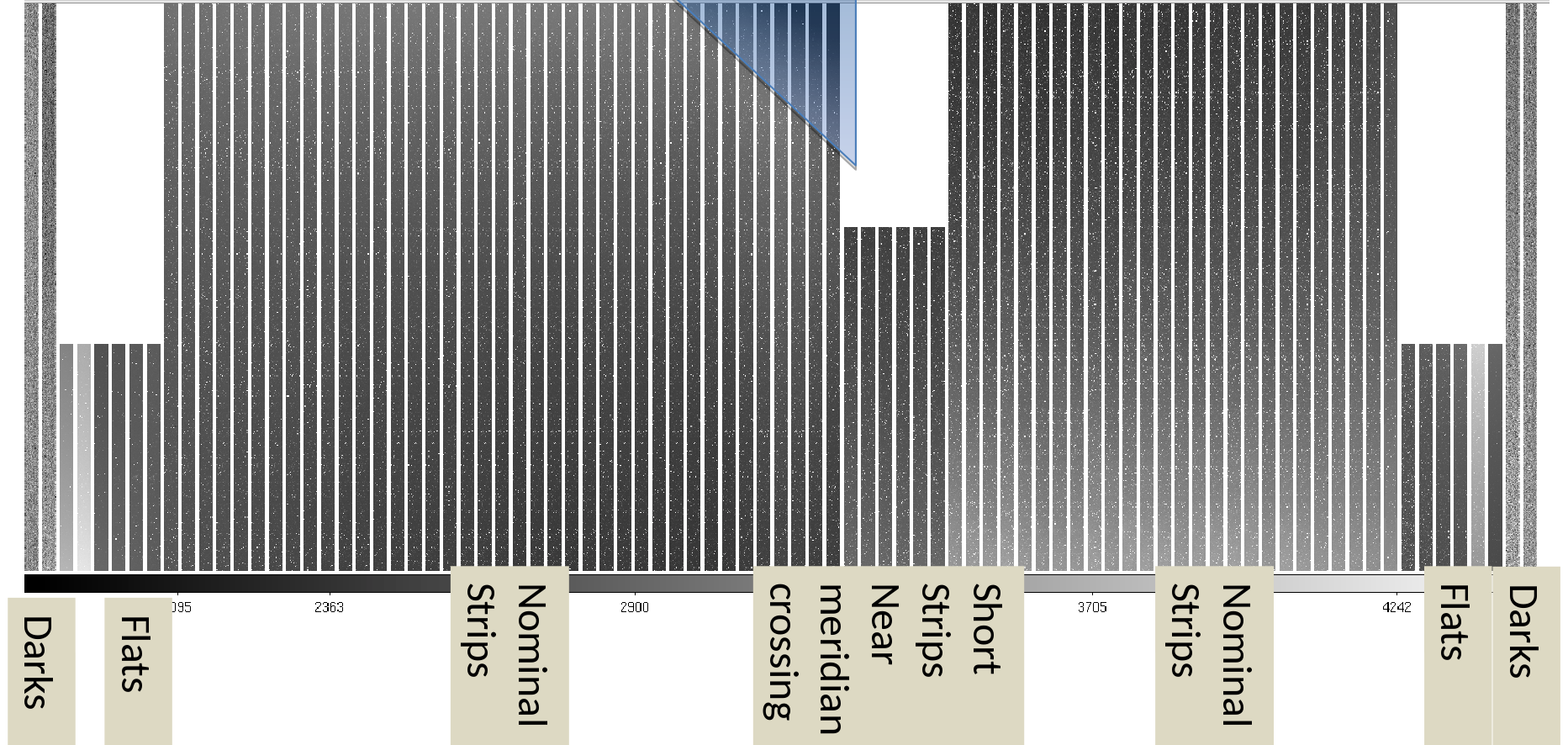}
\caption{Example of observing night with XO. These data were taken by one camera of the unit located at the Observatorio del Teide, Canary Islands, during the night of June 8, 2013. Only a sample of the data is shown. The TDI mode yields long strips instead of square images. The operations are summarized at the bottom from dusk (left) to dawn (right). The blue triangle near the meridian crossing indicates parts of the strips that are contaminated by one of the enclosure's walls (see Section \textit{\nameref{sec: Data selection}}).}
\label{fig: XO observations}
\end{figure}

We observed two strips with respective centers at $RA = 90^{\circ}$ and $RA = 270^{\circ}$ and $Dec$ extending from $+90^{\circ}$ to $+54^{\circ}$. This corresponds to an effective sky area of $520^{\circ 2}$. We observed these strips for twice nine months between 2012 and 2014, one strip at a time depending on their observability. An example of one night observations with XO is shown in Figure \ref{fig: XO observations}. Darks and flats are taken at the beginning and end of the night, science strips are taken the rest of the night. The scan direction is either North or South with a switch occurring around the meridian crossing. Under nominal conditions, we collect from 60 to 100 TDI strips each night depending on the season.

\subsection{Data quality check}

\begin{figure}
\includegraphics[width=\columnwidth]{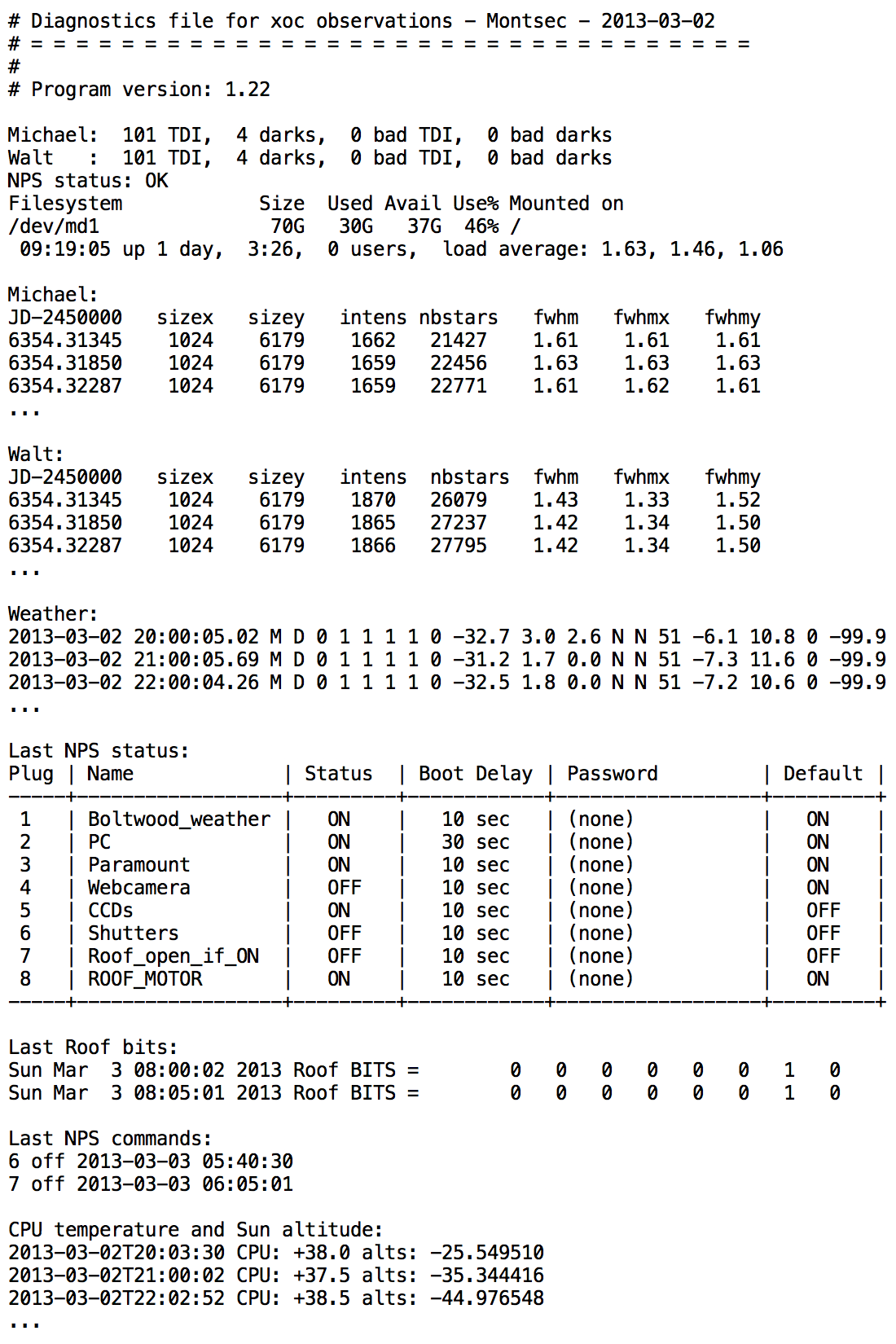}
\caption{Example of nightly diagnostics file for the night of March 2, 2013 for the XO unit at Observatori Astron\`omic del Montsec. ``Michael'' and ``Walt'' are the nicknames of the two cameras. Each camera recorded 101 TDI strips; the diagnostics information for only three strip is shown for clarity (as indicated by `...'). The weather sensor information, NPS state, enclosure's roof state, CPU temperature, and Sun altitude are also reported.}
\label{fig: diagnostics}
\end{figure}

We developed a data quality check software program that analyzes the observations after each night. All science and dark files are listed and their size is checked. Files with a very small size are corrupted (usually they contain a header with no data) and are removed from the list.
Then, for each science image (strip), we measure its size in pixel units, the corresponding number of 1024$\times$1024 sub-images, the median intensity, the sky background, and the number of point sources. We also compute the FWHM in the $x$ and $y$ directions using 100 bright, non-saturated stars. The median intensity and the number of point sources are also computed for each 1024$\times$1024 sub-image. This program is ran for both cameras at each site.
In addition, we summarize the weather data and the CPU temperature over the night. We also report the status of the network power strip (NPS) and of the enclosure roof, the last commands that were issued at the end of the observations, and the available disk space.
The results are sent via email (Figure \ref{fig: diagnostics}) with a separate alert email if the NPS is not in a nominal state. This program runs every day at each site.

\subsection{Data management}

The data are stored on the computers at each site and are transferred periodically on servers at Space Telescope Science Institute (STScI) through internet using a $rsync$ command. A 8 TB storage space was allocated to the project. The full analysis of the data is performed on the STScI servers.

\section{Data reduction}
\label{sec: Data reduction}

The data reduction pipeline is divided in two phases. The first phase consists of reducing the data of each night. The second phase consists of gathering the data taken on the same field over all nights and to build the lightcurves. Data from the six cameras are reduced independently and the lightcurves are merged at the very end of the data reduction.
The pipeline is composed of \textit{IDL} routines that are launched with \textit{Bash} programs. The \textit{Bash} commands iterate on each night and then on each field. The main advantage is that if \textit{IDL} crashes, the \textit{Bash} program just moves on to the next night or field.

\subsection{Carving the strips}

For each night, the strips are carved into $7^{\circ}.2\times7^{\circ}.2$ square images. A World Coordinate System (WCS) solution is found using the astrometry.net software program \citep{Lang2010} plus a six parameter astrometric solution. We use a two step process: first, each square image is centered on the coordinates of a predefined field (listed in Table \ref{tab: xo fields}), then a precise astrometric solution is found. Observing in TDI mode allows us to observe these square fields more efficiently than simply cycling through them using standard telescope pointing and CCD readout, as done by other transit surveys.

\begin{table}
\begin{center}
\caption{Central coordinates of the fields of view observed by XO. Because we observe in TDI mode, fields with the same $RA$ are observed within one scan. Only a small amount of data is obtained for fields 10 to 12.}
\label{tab: xo fields}
\vspace{3mm}
\begin{tabular}{cccccccccccccc}
\hline
\hline
Field & 00 & 01 & 02 & 03 & 04 & 05 & 06 & 07 & 08 & 09 & 10 & 11 & 12  \\
\hline
$RA \rm \; [^{\circ}]$ & 270 & 270 & 270 & 270 & 270 & 90 & 90 & 90 & 90 & 270  & 270 & 90 & 90  \\
$Dec \rm \; [^{\circ}]$ & +90 & +83 & +76 & +69 & +62 & +83 & +76 & +69 & +62 & +55 & +48 & +55 & +48  \\
\hline
\hline
\end{tabular}
\end{center}
\end{table}

\subsection{Image calibration}


About five dark frames and five flat frames are taken for each camera at the beginning and end of the night. They come as strips of the same size as the science images, but because we use the TDI mode, the final darks and flats are 1-D vectors with 1024 elements corresponding to the columns.
For each dark frame, an outlier resistant mean is computed for each column after excluding the first 1024 rows. The resulting 1-D vectors that have a mean value between 500 and 2000 ADU (Analog to Digital Unit) are averaged to create a final dark. Those with higher or lower median values are not used, as they are contaminated by parasitic light for example. One dark is created for each night. In the absence of dark frames, we use the dark of the previous or the following night.

\begin{figure}
\includegraphics[width=\columnwidth]{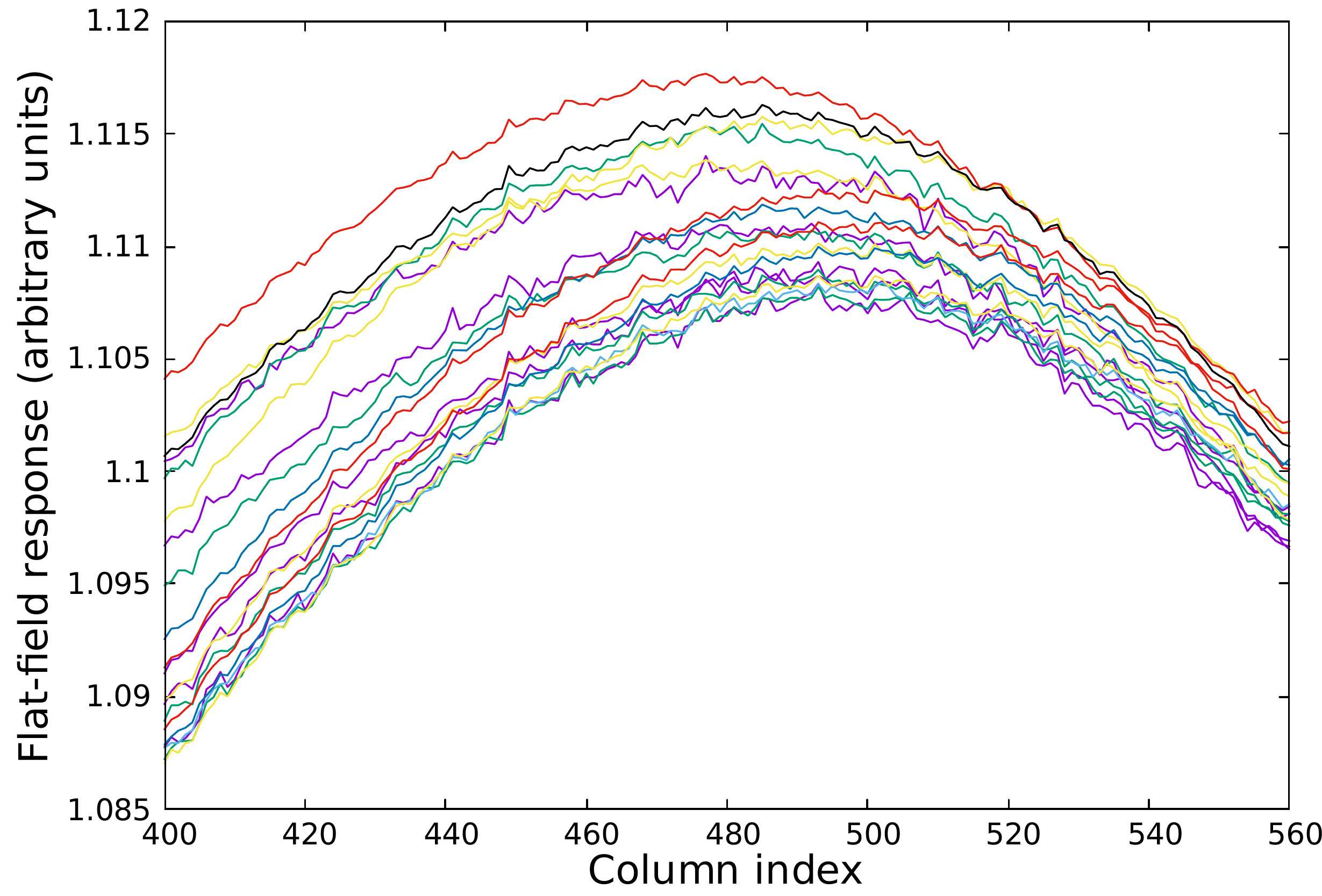}
\caption{Example of flats built from twilight images taken during March, 2013 with one camera, zoomed in between columns 400 and 560. Because we use the TDI mode, flats are one-dimensional vectors with 1024 elements. We normalize and average these flats to built a masterflat for each camera and use it for all the observations.}
\label{fig: flat-fields}
\end{figure}

One flat is built for each camera. We use the TDI images recorded during twilight and calibrate them with the dark. We eliminate pixels that are close to saturation ($>$~50,000 ADU) as well as a region of $3\times3$ pixels around them.
We eliminate stars in the following way. We build a median-filtered image where each pixel is replaced by the median of the $15\times15$ surrounding pixels, and subtract it to the initial image. We compute the outlier-resistant standard deviation of the residual image. Pixels that are above three times this standard deviation are flagged and are removed from the initial image. This yields an image equivalent to the initial image but with the stars removed.
Then, we divide each row by its median in order to give the same weight to all rows. We compute an outlier-resistant mean of the image along the columns (starting at row 1024) and normalize the resulting vector by its median. This yields a flat vector. We average the flat vectors obtained from each twilight image to build a flat vector for each day (Figure \ref{fig: flat-fields}), and we average the daily flat vectors obtained from the first months of observations. 
Permanent warm columns are identified using the darks and are replaced by the average of the 10 neighboring valid columns (here a ``column'' refers to one vector element). We suppress high frequency variations by replacing each element by the average of its 10 neighbors. We normalize this vector by the median of the 50 central elements. This yields a masterflat that we use for all the observations. This procedure is repeated for all six camera.
Finally, the science image calibration is performed by subtracting the dark vector to each row and dividing it by the flat vector.


The strips are affected by warm columns. There are typically 10 warm columns per image except for one CCD that produces about 100 warm columns. About a third are permanent and two thirds vary from image to image. We correct for them in each image in the following way. We compute a smoothed version of each row using a 11-pixel running median and subtract it to the original row. Then, we compute the median of each column in the residual image. Those exceeding a threshold are considered as warm; we use a threshold of 15 ADU for the six CCDs after a visual inspection. For these columns, this median is subtracted to the same column in the original image in order to remove the excess counts. We record the number of corrected columns, their indices, and their original values in the image header.

\subsection{Flux extraction}

\begin{figure}
\includegraphics[width=\columnwidth]{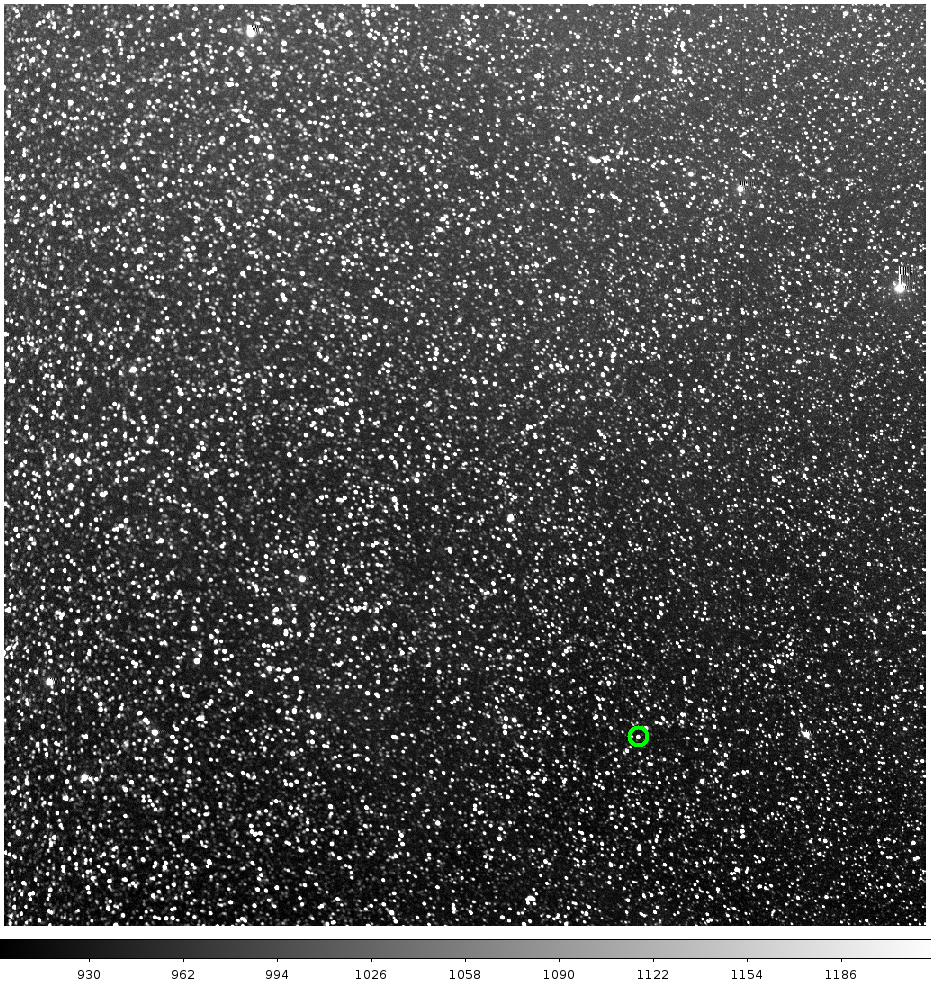}
\caption{Reference image for field 06. Point sources identified in all the images of this field are matched against this image. The image size is 1024$\times$1024 pixels and $7^{\circ}.2\times7^{\circ}.2$ on the sky. The gray scale at the bottom indicates the intensity in ADU (most bright pixels are beyond the scale limit). The star hosting the transiting hot Jupiter XO-6b is indicated by a green circle.}
\label{fig: reference image}
\end{figure}

\begin{figure}
\includegraphics[width=5.7cm]{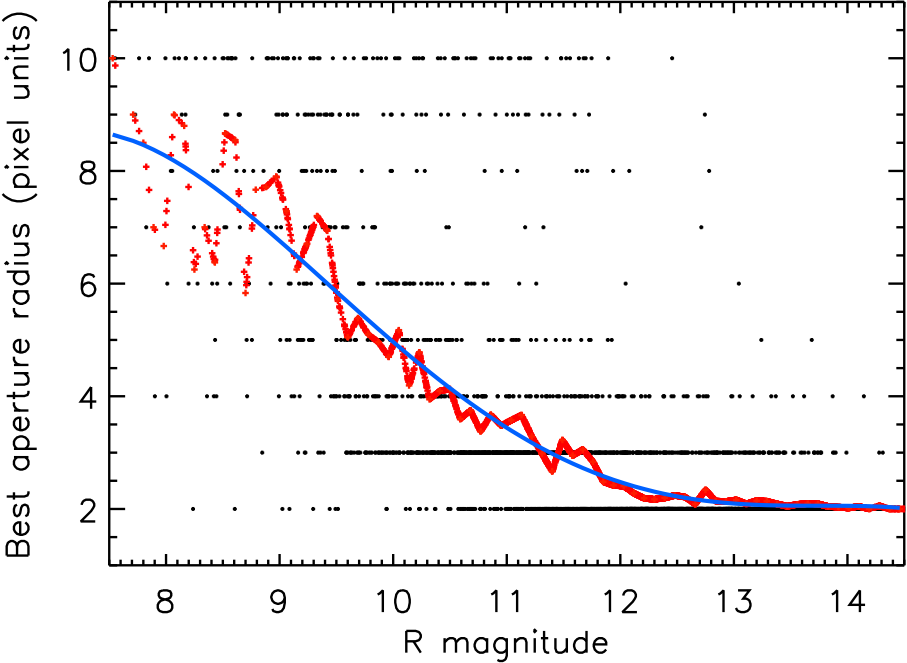}
\hfill
\includegraphics[width=5.7cm]{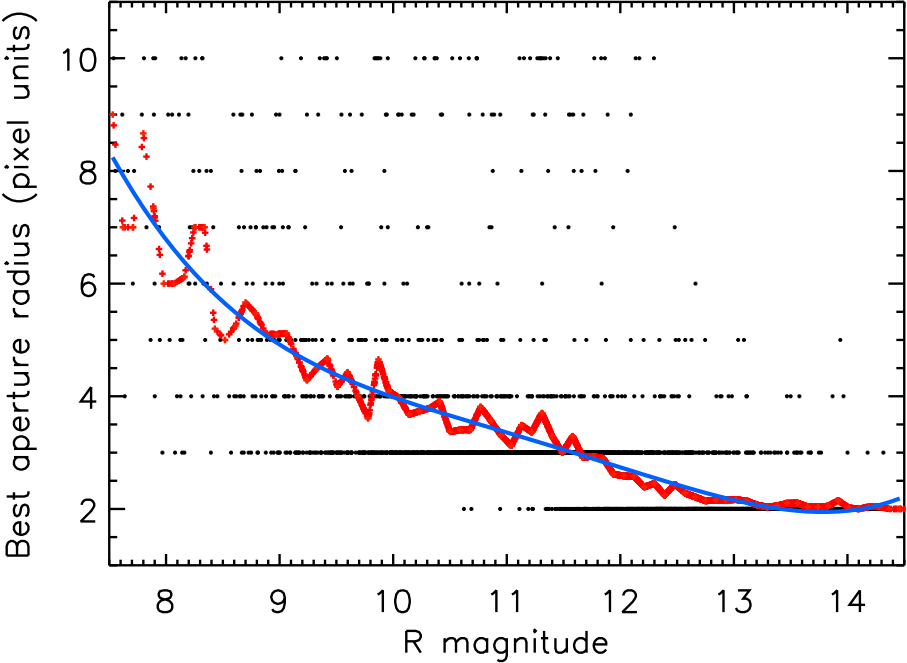}
\caption{Best photometric aperture as a function of stellar $R$ magnitude for two cameras, at Vermillion Cliffs Observatory (left) and at Observatori Astron\`omic del Montsec (right). Each star is represented by a black dot. A mean is performed in each magnitude interval (red points) and a fourth order polynomial is fitted to define an aperture function (blue line). We use this function to define an aperture for each star.}
\label{fig: aperture function}
\end{figure}

We select around 6000 target stars in each field for the photometry using an image taken under excellent conditions (the ``reference image'', Figure \ref{fig: reference image}). Stellar fluxes are extracted by circular aperture photometry using the \textit{Stellar Photometry Software} program \citep[\textit{SPS,}][]{Janes1993}. We compute the sky background for each star using an annulus around the aperture and subtract it. To measure the flux, \textit{SPS} does not simply sum the pixels inside the aperture: instead, it computes an intensity profile by interpolating between pixels, and measures the flux by integrating this profile over the aperture size. In the original version of the XO pipeline, only one aperture of 3 pixel radius was used for all the stars. We improved this by optimizing the aperture for each star, as described next. For this process, we use only high quality data that were selected after a first flagging procedure and visual inspection of the star -- epoch magnitude arrays (see Sections \textit{\nameref{sec: Building the lightcurves}} and \textit{\nameref{sec: Data selection}}). 
For each star, we build lightcurves for nine apertures with radii ranging from 2 to 10 pixels, calculate their RMS, and select the aperture that leads to the smallest RMS. We use the point-to-point RMS rather than the standard RMS so the aperture optimization is not affected by long term trends. This is justified because point sources are identified in each image and the apertures are placed at their centroids.
The fields of view are crowded so the best aperture for a given star may be affected by other sources, and it shows large fluctuations for stars of similar magnitudes. To suppress these fluctuations, we build a function that sets the optimal aperture for each stellar magnitude. We divide the magnitude range [6, 15] into 100 bins, calculate the mean best aperture in each bin, and fit this distribution by a fourth order polynomial (Figure \ref{fig: aperture function}). The last step is an iterative process to have the magnitude of each star correspond to that measured in its optimal aperture. 
We run this process for two representative fields (00 and 02) and average them to obtain a final aperture function. We build an aperture function for each camera. Differences between cameras are due to different focus or stability, for example. Finally, the optimal aperture for each star is given by the aperture function rounded to the closest integer (because only a small number of apertures can be used, due to a limitation of \textit{SPS}).

\subsection{Building the lightcurves}
\label{sec: Building the lightcurves}

The strip carving, astrometry, image calibration, and flux extraction are performed for each night. We obtain a star coordinate -- magnitude file for each image. Because the \textit{SPS} software identifies point sources in each image before doing the photometry (it does not handle input coordinates), we must correlate the sources found in each image to build the lightcurves. This correlation is made against the reference image. We include an image distortion correction performed in $x, y$ rather than in $RA, Dec$ because the observed fields are close to the celestial pole which yields degeneracies in $RA$. Then, we build a star -- epoch array that gathers the magnitudes recorded for all stars at all epochs. Similar star -- epoch arrays are built for the magnitude error, $RA$, $Dec$, $x$, $y$, sky background, airmass, and PSF cross-correlation parameter. We obtain one set of such arrays for each night, field, and camera.

The second phase of the pipeline is ran for each camera and each field. We combine all the nights together: we correlate the star coordinates from night to night and build one set of star -- epoch arrays covering all the observations. Examples of such arrays are shown in Figure \ref{fig: cal arrays}. Because we are interested in bright stars, we truncate these arrays and keep only the 2000 brightest stars in each field, which corresponds to a limit $R$ magnitude around 12.5 and a photometric precision of 2-3\%. We use only these stars in the rest of the pipeline.

\subsection{Photometric calibrations}

The next steps are photometric calibrations that are essential to improve the lightcurves from the raw photometry. These calibrations are applied independently for each scan direction (the strips are observed scanning North or South which yield different systematic effects). First, we compute the mean magnitude of each lightcurve and subtract it. From this point, the lightcurves consist of residual magnitudes rather than absolute magnitudes and we work with the residual magnitude arrays. We select reference stars in the following way. For a given star $s$, we subtract the photometric time series of $s$ from those of all the stars and evaluate the mean absolute deviation (MAD) of the resulting time series. Stars are sorted by increasing MAD; the first index is excluded because it corresponds to star $s$, and the following $N$ stars are kept as reference stars. Then, we build a reference lightcurve using an outlier-resistant mean of the $N$ reference stars, and subtract it to the lightcurve of star $s$. This yields a calibrated lightcurve for star $s$. We use $N = 10$. Lightcurves of stars with a bright nearby reference star are usually of better quality. For example, XO-6 has a reference star of magnitude $R = 9.6$ located at 5.6 arcmin (14 pixels) separation. This reference star makes XO-6b a good target for atmospheric characterization by differential spectrophotometry from the ground, if both stars can be placed on the detector. 
Then, for each epoch, we remove a 3rd order polynomial corresponding to low-frequency variations of the magnitude residuals in the CCD's \textit{x,y} plane, also called ``L-flats". We also remove a linear dependence of each lightcurve with airmass. At the end of the process, the mean magnitude of each lightcurve is subtracted again.

\subsection{Data selection}
\label{sec: Data selection}

\begin{figure}
\includegraphics[width=\columnwidth]{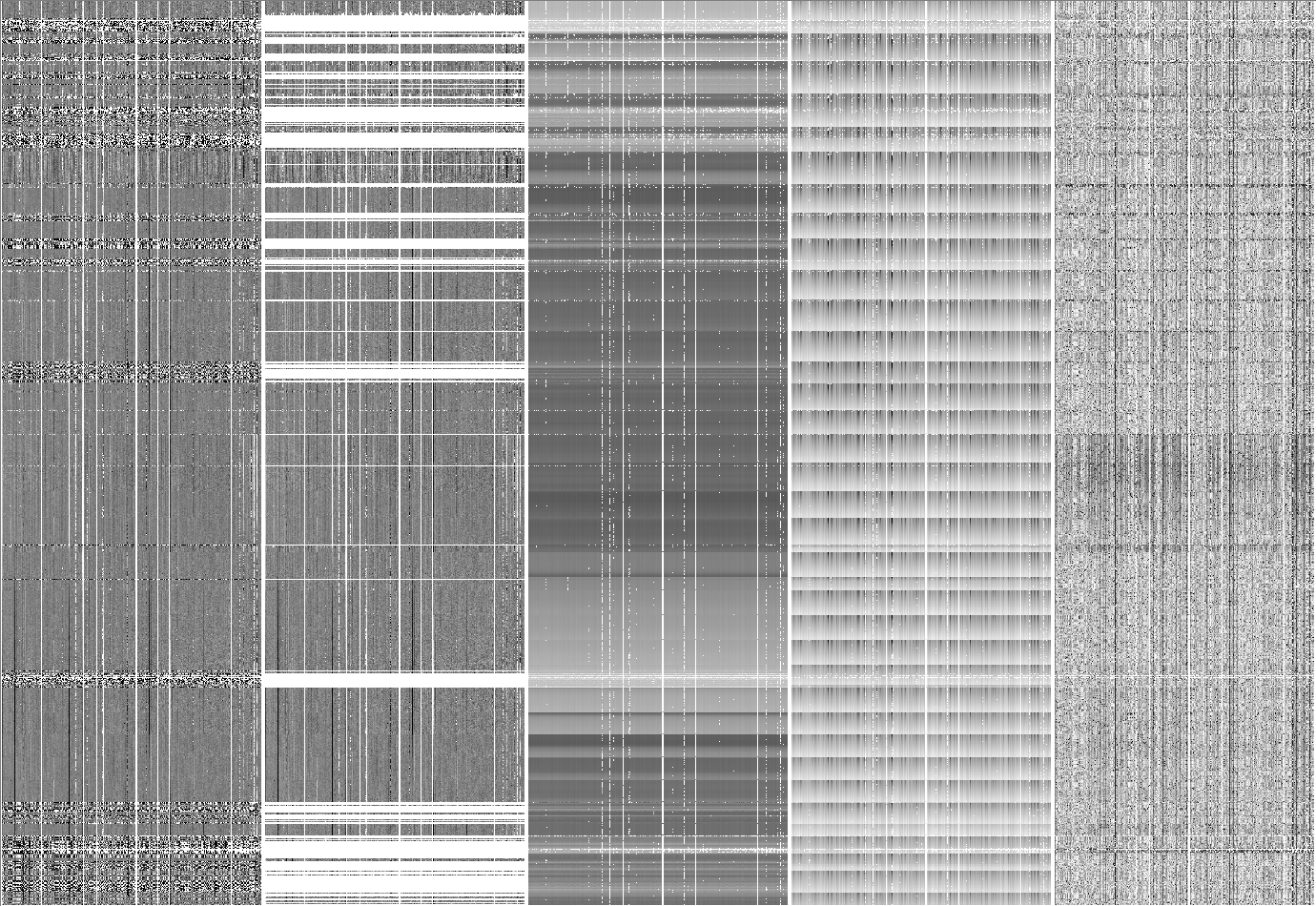}
\caption{Star -- epoch data arrays generated by the XO pipeline. Stars and epochs are sorted on the $x$ and $y$ axis by increasing magnitude and time, respectively. The full arrays contain 2000 stars and around 12,000 epochs. A subset of 475 stars $\times$ 1640 epochs is displayed here, and corresponds to field 02 observed with one camera at the Observatorio del Teide between March 12 and April 25, 2013. From left to right: magnitude residuals before flagging, magnitude residuals after flagging, sky background, airmass, cross-correlation function. In the magnitude residual arrays, stars that have been discarded appear as white columns, variable stars appear as alternatively dark and bright columns, and epochs with an increased dispersion $\sigma_{ep}$ appear as black and white regions (before they are flagged).}
\label{fig: cal arrays}
\end{figure}

Parts of the data are affected by poor weather, high sky background, instrumental malfunctions, etc., and must be removed. A lot of effort is put into inspecting the magnitude arrays, identifying the low quality parts, searching for the causes, defining criteria and procedures to flag them, and adjusting the flagging parameters. We summarize below the main causes of low quality data and the data selection procedures. Figure \ref{fig: cal arrays} shows an example of star -- epoch magnitude arrays before and after the flagging.

\begin{figure}
\includegraphics[width=\columnwidth]{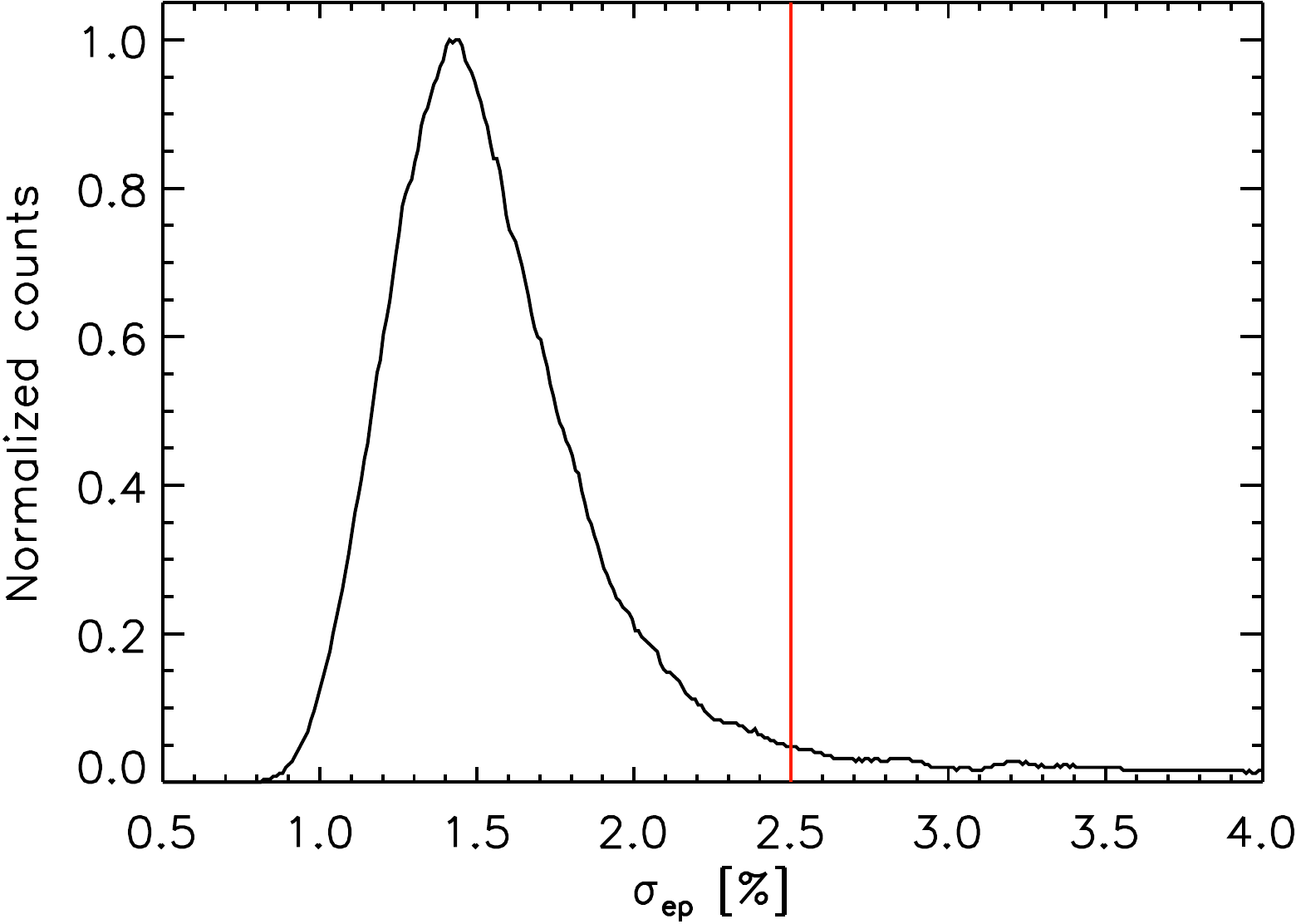}
\caption{Histogram of the standard deviation of the magnitude residuals at each epoch, $\sigma_{ep}$, for field 00 observed with one camera at Vermillion Cliffs Observatory over the whole duration of the observations (black line). The tail of the histogram corresponds to data taken under poor conditions, generally due to bad weather, that yield larger values of $\sigma_{ep}$. A threshold (red line) is set manually for each field and each camera to flag these data. A similar approach is used for other flagging parameters.}
\label{fig: histo sigep}
\end{figure}

\begin{itemize}

 \item The main cause of bad data is poor weather. To quantify this, we calculate the standard deviation of the magnitude residuals at each epoch, $\sigma_{ep}$, using stars in the magnitude range [8.3, 10]. We flag epochs with $\sigma_{ep}$ above a given threshold that we set manually for each camera and each field (Figure ~\ref{fig: histo sigep}). These thresholds are between 1.9\% and 4.5\% with an average at 2.3\%.
 
 \item We flag data with a sky background larger than 10,000 ADU and those with an airmass larger than 3.
 
 \item We flag epochs where one of the enclosure's wall is visible in the images, which occurs when the observed field is at low altitude. Although part of the image could be used, some reference stars may be missing so we discard the whole image. These images show a large difference of sky background between the part pointing at the sky and that pointing at the wall. We identify them by calculating the standard deviation of the sky background across the image normalized to its mean. Those with a value larger than a given threshold are flagged.
 
 \item We flag epochs where less than half of the stars have a valid magnitude measurement. 

 \item We flag epochs where the cross-correlation parameter averaged over all stars is lower than a threshold value, ranging from 0.61 to 0.74 depending on the camera and field. This parameter is a crude measurement of the PSF shape. In particular, it identifies images that are affected by imperfect tracking, which have a low correlation parameter (see Section \textit{\nameref{sec: Mount stability}}).

\end{itemize}

\subsection{Correction of systematic effects and final lightcurves}

We correct the remaining systematic effects in the cleaned residual magnitude arrays using the \textit{SysRem} algorithm \citep[\textit{Systematic Removal},][]{Tamuz2005}. We correct for 10 eigenfunctions using 20 iterations, still per camera and field independently. Finally, we combine the lightcurves and other parameters obtained from the six cameras, interleaving the epochs and sorting them by ascending Julian date. This yields one set of star -- epoch arrays for each field.

\section{Instrumental performances}
\label{sec: Instrumental performances}

In this section, we review some technical issues that we encountered along the observations and present the photometric performances of the instruments.

\subsection{FWHM variations}
\label{sec: FWHM variations}

\begin{figure}
\includegraphics[width=5.7cm]{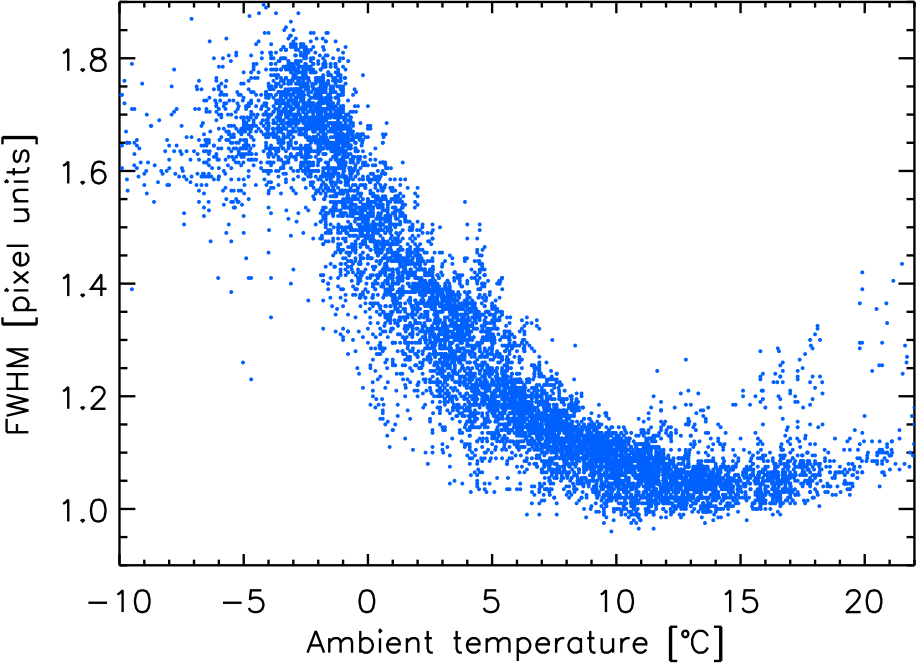}\hfill
\includegraphics[width=5.7cm]{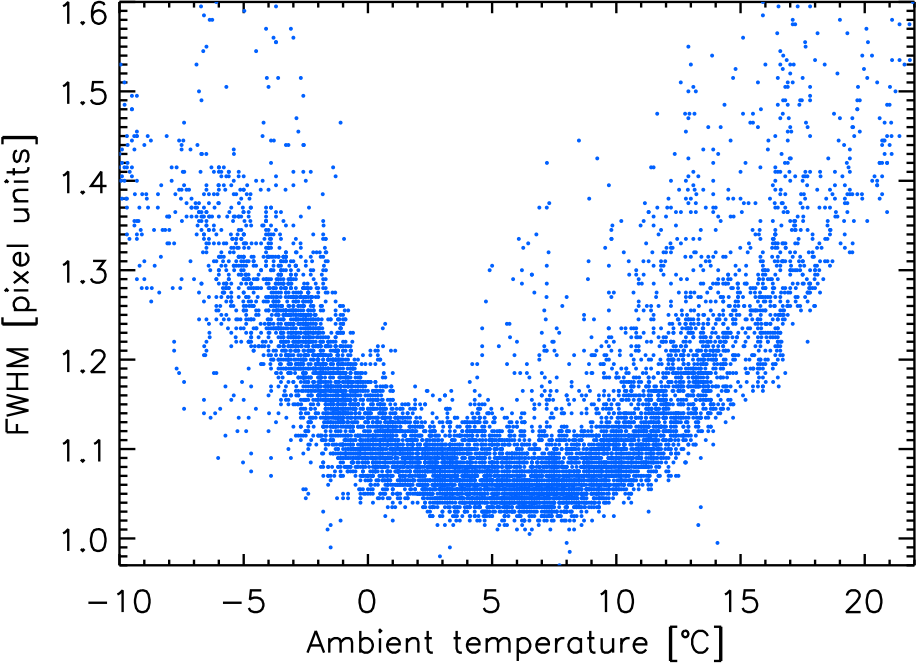}
\includegraphics[width=5.7cm]{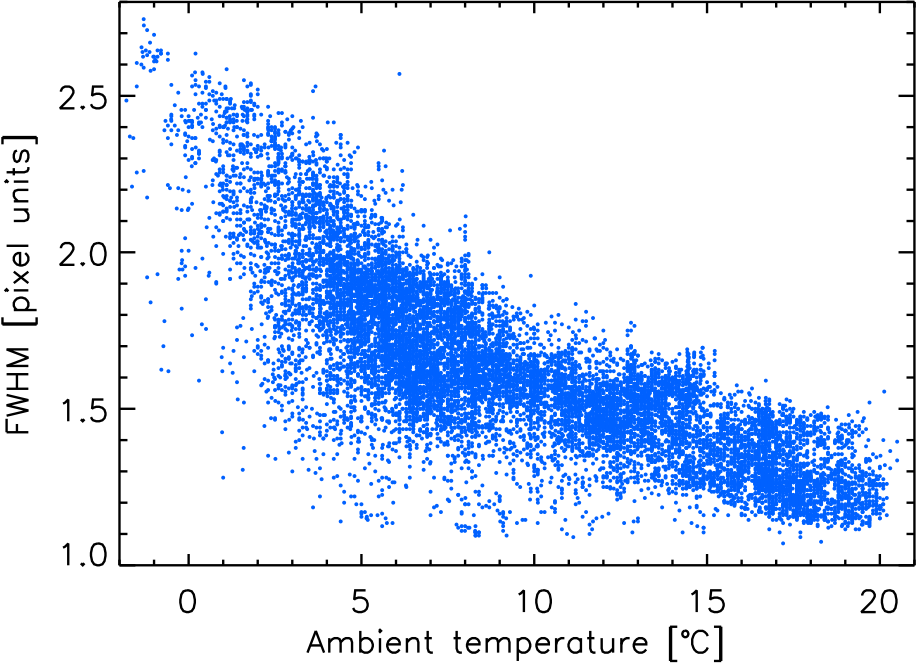}\hfill
\includegraphics[width=5.7cm]{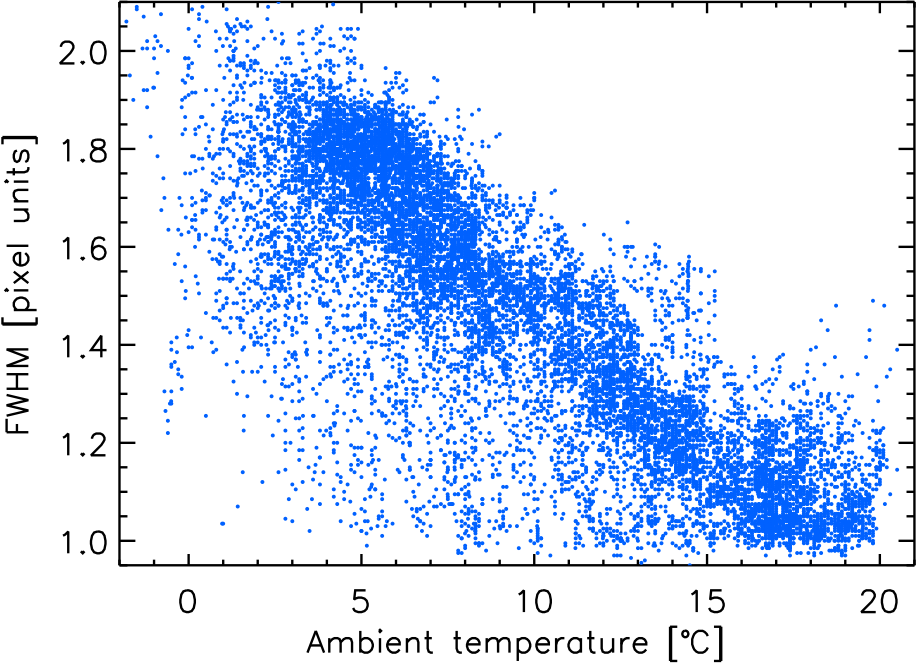}
\includegraphics[width=5.7cm]{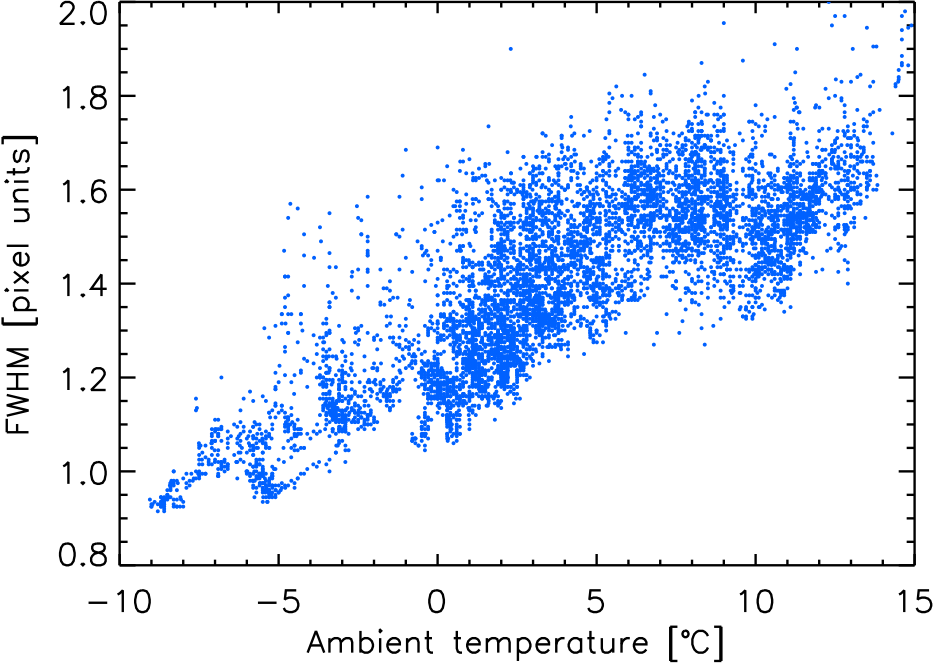}\hfill
\includegraphics[width=5.7cm]{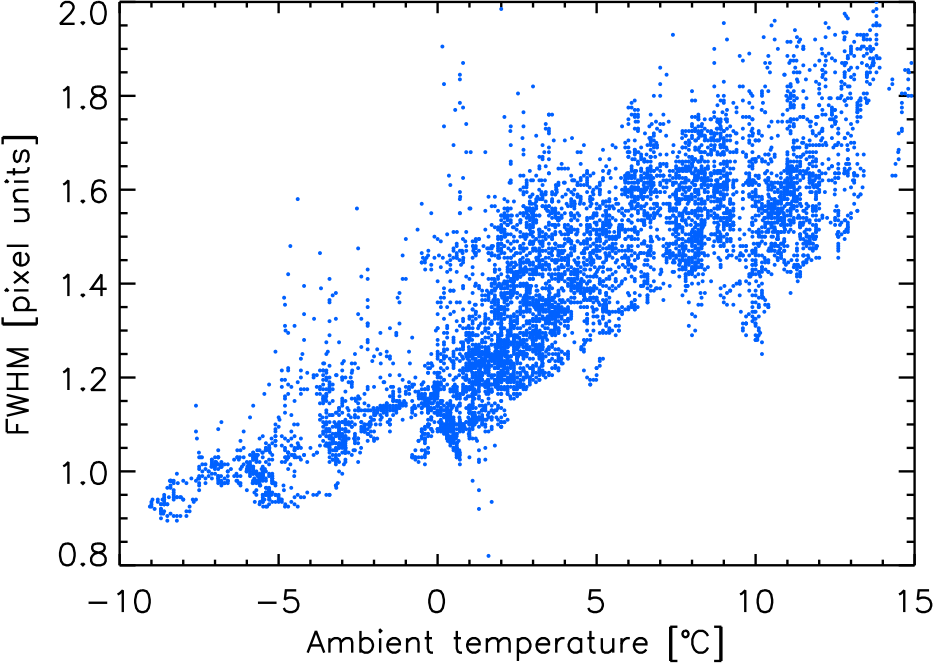}
\caption{Variations of the PSF FWHM as a function of the ambient temperature during the first months of observations for the six cameras. Top: Vermillion Cliffs Observatory; middle: Observatorio del Teide; bottom: Observatori Astron\`omic del Montsec. These variations were minimized afterwards by improving the temperature control system.}
\label{fig: fwhm variations}
\end{figure}

After analyzing the first months of observations, we found significant variations of the PSF FWHM that are well correlated with the ambient temperature (Figure \ref{fig: fwhm variations}). This effect is common and due to the lenses. Such PSF variations affect the flux measurements. To minimize this effect, the lenses were surrounded by a thermal regulation system made of flat and soft resistor strips for heating, a thermal probe, and a synthetic isolator, but this system was not efficient enough. After the first nine months, we increased the heating power and the isolation layer which reduced these FWHM fluctuations.

\subsection{Mount stability}
\label{sec: Mount stability}

\begin{figure}
\includegraphics[width=\columnwidth]{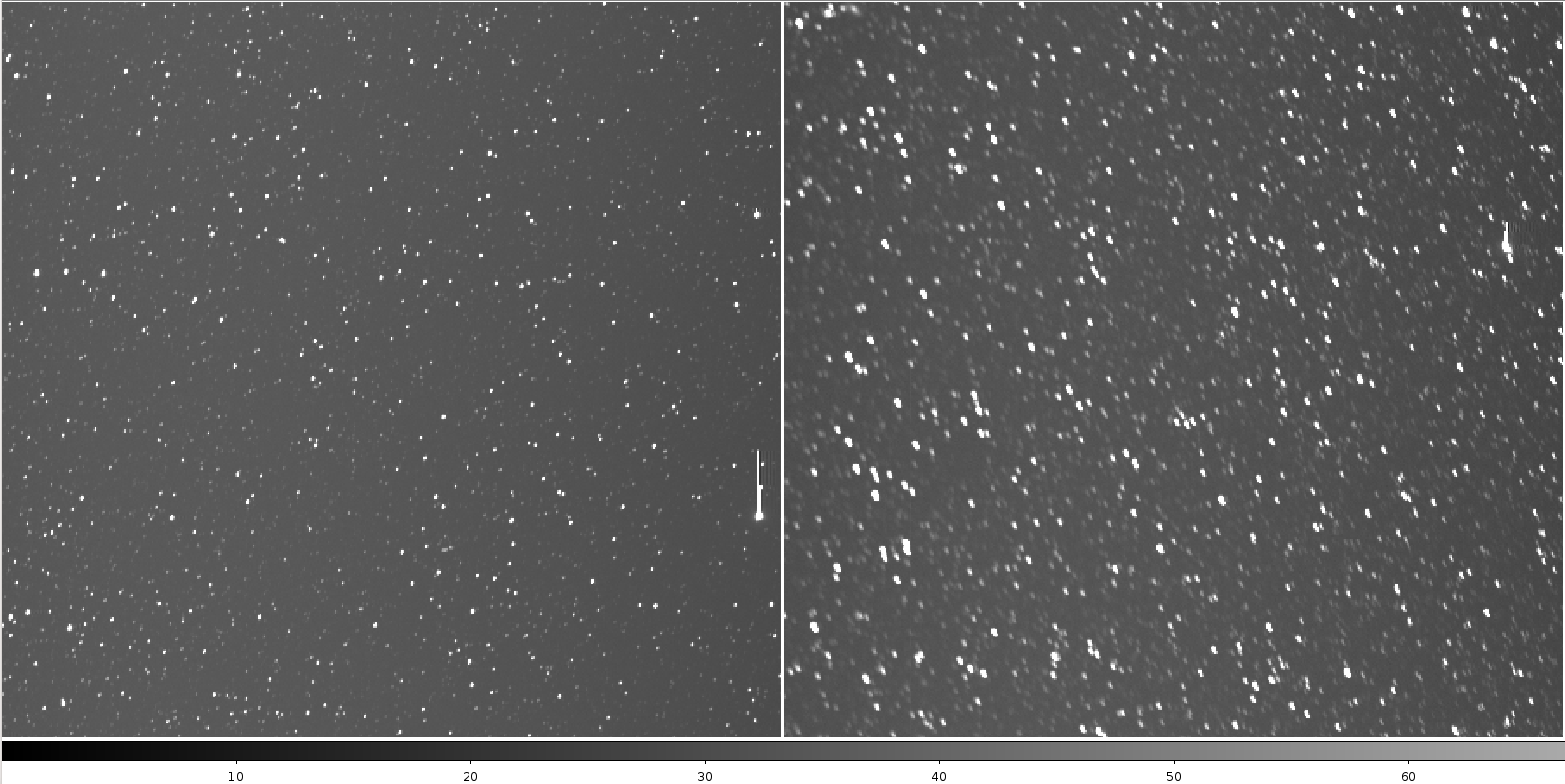}
\caption{Zoom on a standard image (left) and on one with elongated PSFs due to a mount defect (right). This behavior is seen only for a small fraction of the data taken at Observatorio del Teide.}
\label{fig: elongated fwhm}
\end{figure}

Some series of images show elongated PSFs or even two point sources separated by a few pixels instead of a single star (Figure \ref{fig: elongated fwhm}). The cause is a mechanical defect in the mount resulting in an imperfect tracking. This behavior is seen only for a small fraction of the data taken at Observatorio del Teide and was corrected after the first nine months of observations. The affected images are discarded following a procedure described in Section \textit{\nameref{sec: Data selection}}.

\subsection{Interferences}

\begin{figure}
\includegraphics[width=6.15cm]{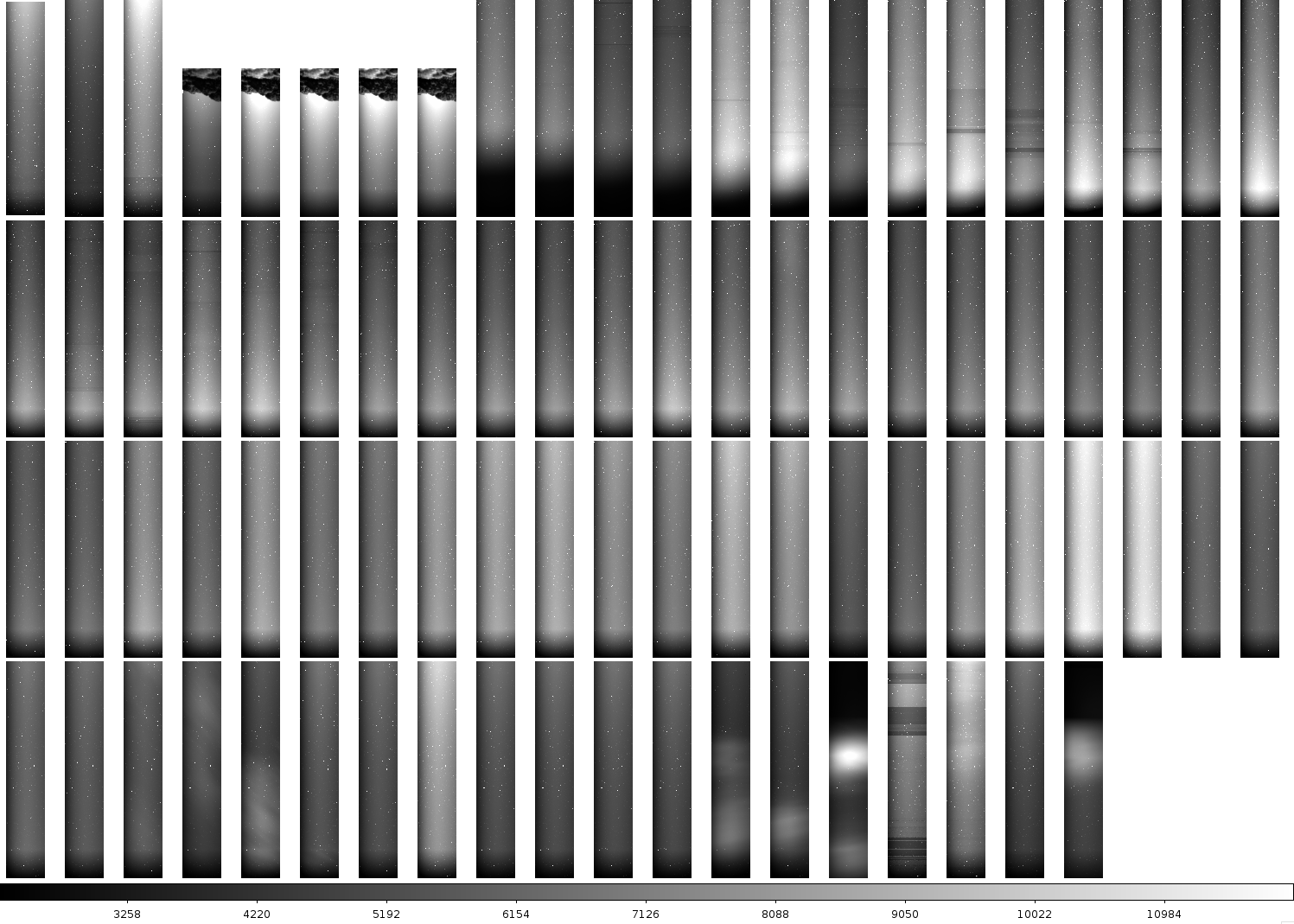}\hfill
\includegraphics[width=5.4cm]{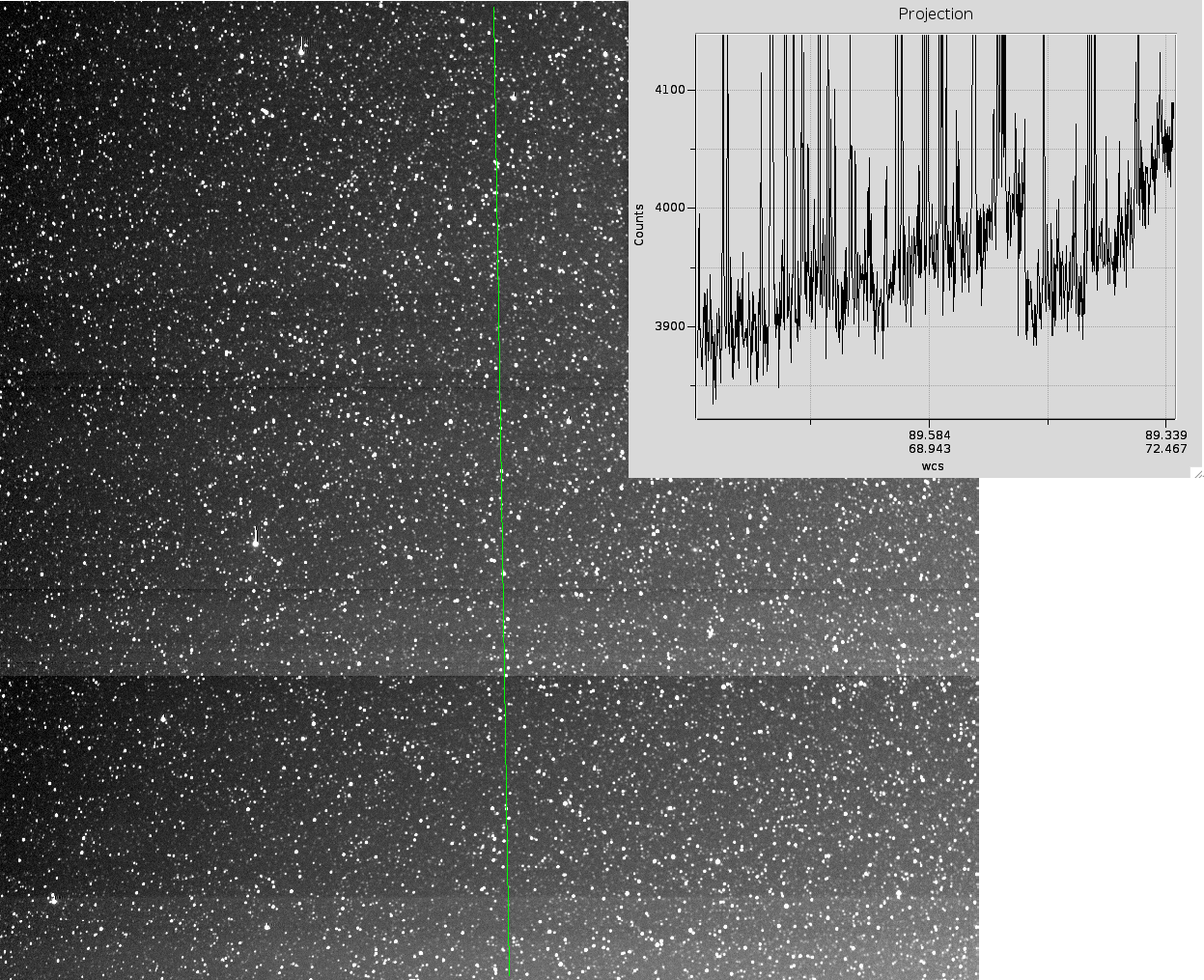}
\caption{Example of images affected by electrical interferences at the Observatorio del Teide. Left: effect on the strips taken during the night of February 25, 2013. Some strips are affected at the beginning and end of the night. The scale in ADU is indicated at the bottom. Right: effect on a square image (obtained after carving the parent strip) and projection along the $y$ direction downwards.}
\label{fig: interferences}
\end{figure}

Another defect is the presence of electrical interferences that create divisions within a strip, each with a specific background level, usually separated by sharp horizontal boundaries, and which vary from one strip to another (Figure \ref{fig: interferences}). This effect is seen for a small fraction of the data taken at Observatorio del Teide and for a very small fraction of the data taken at Vermillion Cliffs Observatory. We solved this effect by improving the electrical isolation of the systems. No clear consequence is identified on the photometry probably because the excess counts are removed while subtracting the sky background, so we keep these data as they are.

\subsection{Photometric precision}
\label{sec: Photometric precision}

\begin{figure}
\includegraphics[width=\columnwidth]{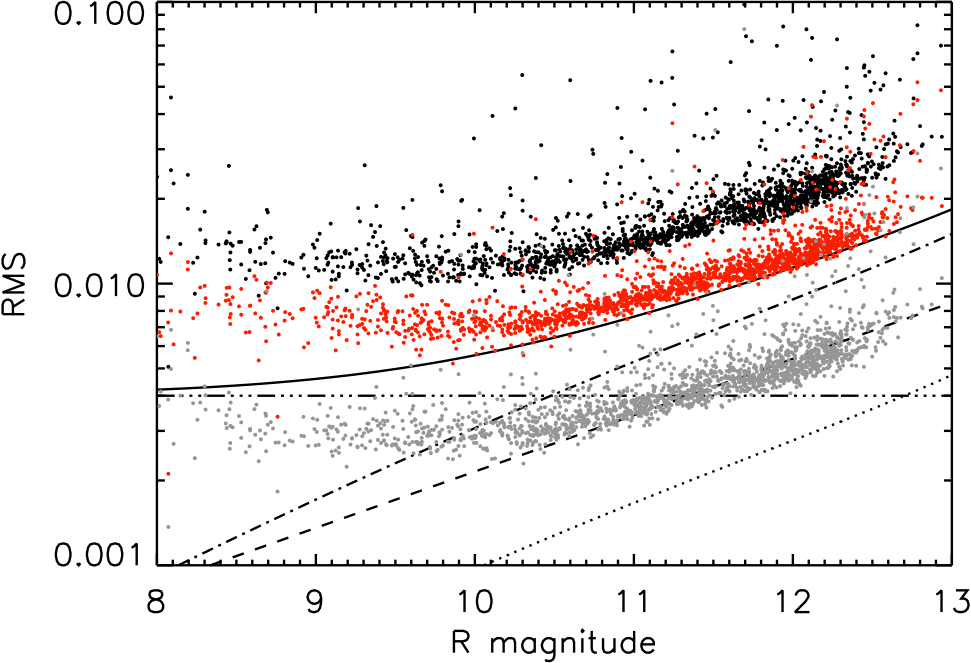}
\caption{RMS -- magnitude diagram for the stars of field 00 obtained from one camera at Vermillion Cliffs Observatory over the full duration of the observations (twice nine months). Each point represents the standard deviation of the lightcurve as a function of the stellar $R$ magnitude. The true RMS (black dots) and the point-to-point RMS (red dots) are calculated from the full unbinned lightcurves with one point taken every 6.3 minutes on average. We also show the point-to-point RMS on a one-hour timescale obtained after binning the lightcurves with a 10-point boxcar (gray dots). Several noise components computed theoretically for the unbinned lightcurves are indicated: Poisson noise (dashed line), sky background noise (dash-dot line), read-out noise (dotted line), a systematic noise floor arbitrarily set at 0.4\% (dash-dot-dot line), and total noise (plain line).}
\label{fig: RMS diagram}
\end{figure}

To evaluate the photometric precision of the instruments, we calculate the standard deviation of the lightcurves over all the observations and report them in a RMS -- magnitude diagram. An example is shown in Figure \ref{fig: RMS diagram} for one camera and one field. We obtain a precision between 1\% and 3\% for stars of magnitude 9 to 12.5 considering all the exposures, on a timescale of 6.3 minutes. This precision is twice larger than the theoretical prediction, on average. We also calculate the point-to-point RMS which does not consider correlated noise. The point-to-point RMS agrees well with the theoretical predictions for stars fainter than magnitude 11 and is slightly larger for brighter stars. The difference between the true and point-to-point RMS indicates that lightcurves are dominated by correlated noise, which is often the case for ground-based time-series photometric observations. For XO, the large amount of collected data by six different cameras from three different sites on the same fields averages these variations and reduces the RMS drastically when folding the lightcurves at given periods, as shown in Figures \ref{fig: xo poster}, \ref{fig: lightcurves}, and \ref{fig: lightcurves long}. For example, after folding the lightcurve of the hot Jupiter XO-6b at the orbital period of the planet, the RMS is 1.3 mmag on a 30 min timescale (the ingress and egress duration) and 0.8 mmag on a 1.5 hour timescale (half the transit duration).

\section{Planet search}
\label{sec: Planet search}

In this section, we present the search for transiting exoplanets from the extracted lightcurves. Several steps are necessary: searching for transit signals, selecting viable planet candidates, and confirming or rejecting them through follow-up observations. Then, we compare the number of detected planets with the expected yield.

\subsection{Search for periodic signals}
\label{sec: Search for periodic signals}

We search for periodic signals in the lightcurves using the BLS algorithm \citep[Box Least Square,][]{Kovacs2002}. This program searches for square shapes in the lightcurves representing transit-like events. Because we are interested in short and long period planets, we perform the search over a wide period range: $0.4 < P < 100$ days. The frequency comb is set by several parameters: the expected transit duration $\tau$, the total extent of the observations $t_{obs}$ (two years), and the maximum number of lightcurve bins $N_{bin}$ that BLS can handle. The transit duration depends on $P$ but also on the stellar density $\rho_\star$. We use the known exoplanet population discovered by the ground-based transit surveys WASP and HAT to bracket the range of stellar densities and adopt a range $0.15 < \rho_\star < 5 \rm \; g\,cm^{-3}$. This bracketing is necessary to limit the search to plausible systems.

\begin{figure}
\includegraphics[width=\columnwidth]{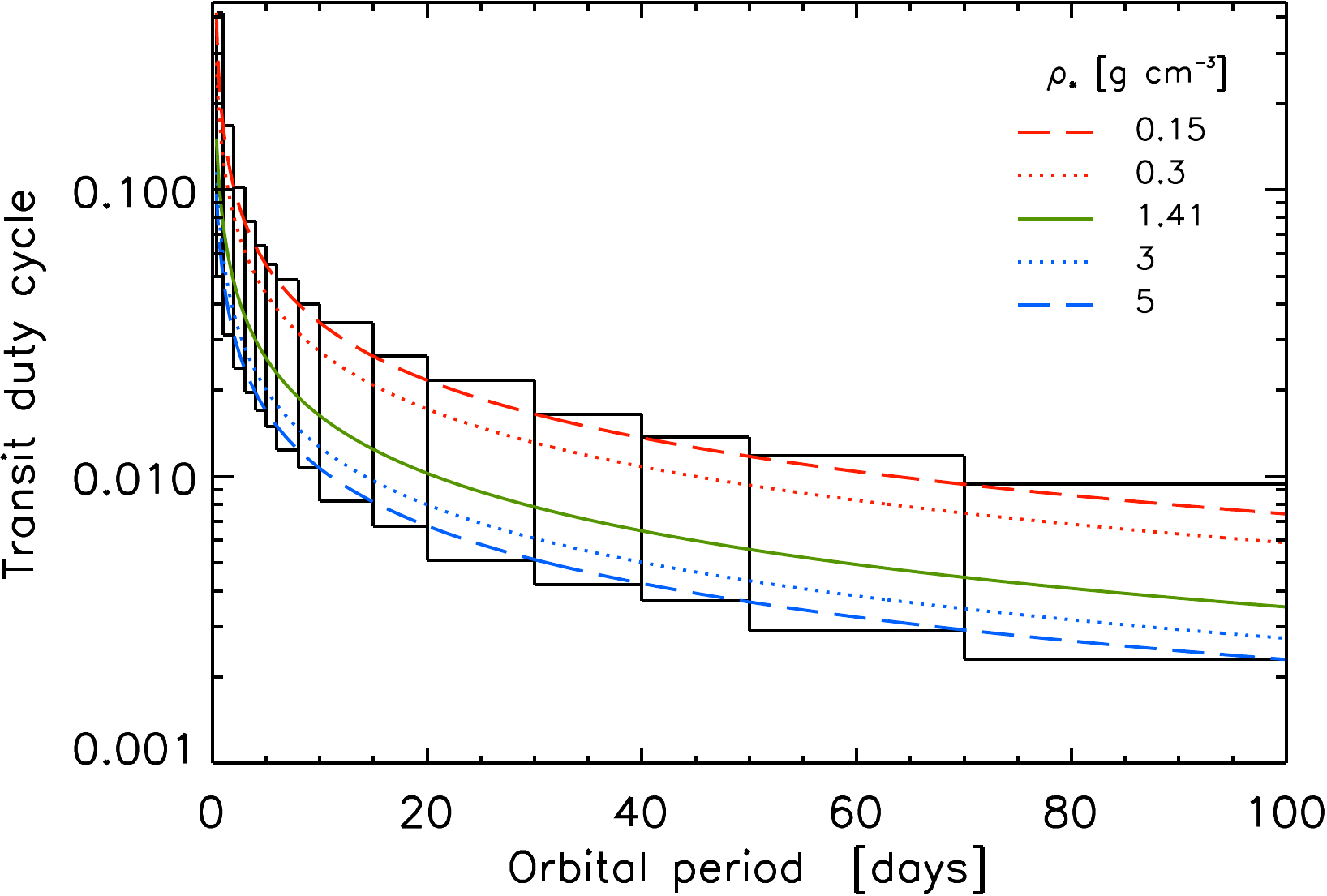}
\caption{Transit duty cycle $q$ as a function of orbital period $P$ for different stellar densities $\rho_\star$. Values of $\rho_\star$ from top to bottom are 0.15, 0.3, 1.41 (Sun), 3, and 5 $\rm g\,cm^{-3}$ and are represented by a red dashed, red dotted, green plain, blue dotted, and blue dashed line respectively. We split the period search range into 15 intervals and define the allowed duty cycles in each of them using the extremum values obtained for the stellar densities 0.15 and 5 $\rm g\,cm^{-3}$, as indicated by the boxes. A specific frequency comb is built for each interval.}
\label{fig: transit duty cycle}
\end{figure}

The transit duration $\tau$ varies widely within the period search range so a unique frequency comb is not appropriate (again it would yield too many unphysical detections). We divide the period range into 15 intervals with sizes following approximately a logarithmic scale. For each interval, we calculate the limits of the plausible transit duty cycles $q$, where $q\, = \, \tau / P$, and use them to build a specific frequency comb (Figure \ref{fig: transit duty cycle}).
Each frequency comb has a constant spacing in $\Delta f / f$, which we set to keep a maximum timing error over $t_{obs}$ as one fourth of the transit duration. We also set the number of lightcurve bins to keep a minimum of four in-transit bins. This ensures that the transit events overlap well with each other when folding the lightcurve. We run the BLS search in each interval separately. During this search, we refine the frequency comb with nine times more frequencies which we apply locally, first around the forty best frequencies, then around the best frequency. For each lightcurve, BLS returns the period that yields the largest power. In addition, we calculate several parameters corresponding to that signal: $q$, $\varphi_{min}$ and $\varphi_{max}$ (the minimum and maximum in-transit phases), the transit depth $\delta$, and $\alpha$. This latter parameter is equivalent to the signal to noise ratio of the transit: $\alpha = \delta / \sigma \times \sqrt{N \times q}$, where $\sigma$ is the standard deviation of the residual lightcurve after subtracting the transit signal, $N$ is the number of data points, and $N \times q$ is the number of in-transit data points (assuming that the points are equally distributed).
 
We run BLS twice. After the first run, the histogram of identified periods shows several peaks that are mostly centered around integer values below 15 days, which indicate aliases of the day-night cycle. Another peak around 29 days is present and is probably due to the Moon cycle. We define filters manually to exclude periods around these peaks and perform a second run.

\begin{figure}
\includegraphics[width=\columnwidth]{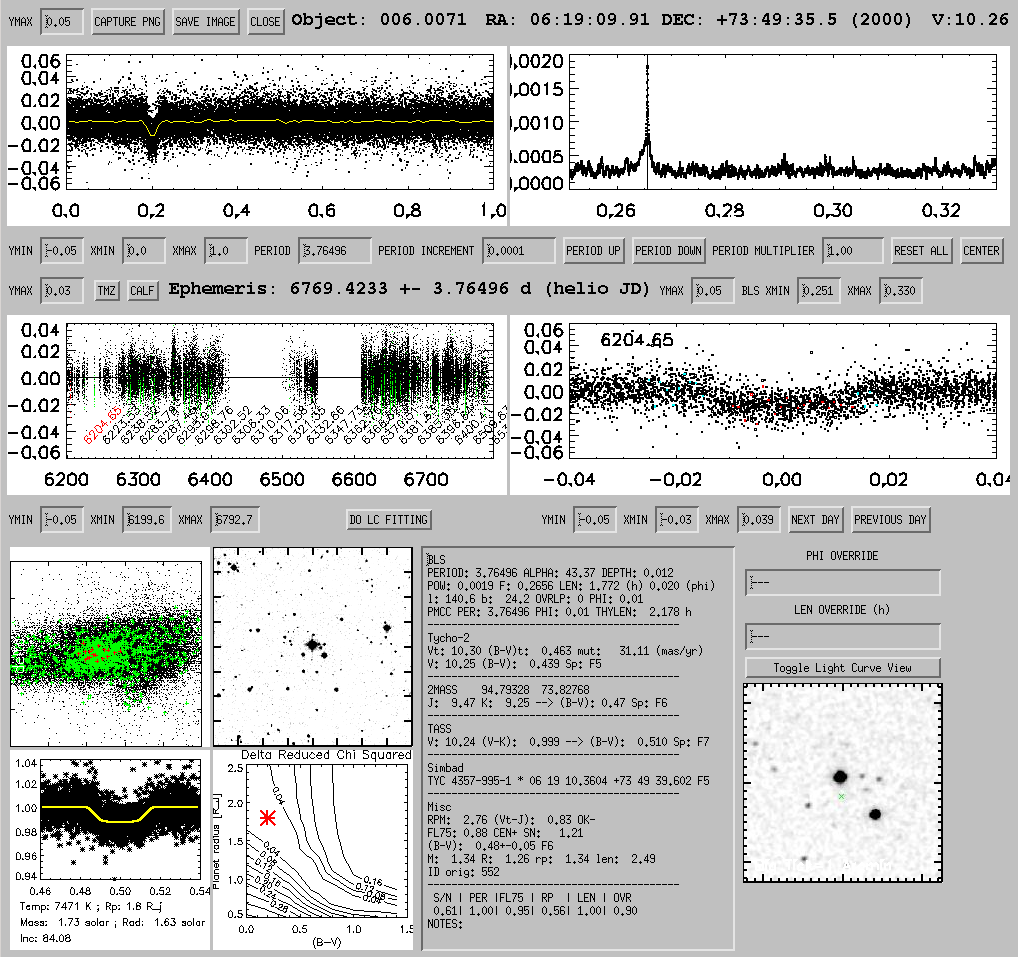}
\caption{Image of the interactive interface that is used to analyze lightcurves and identify transit candidates. This example is that of the hot Jupiter XO-6b \citep{Crouzet2017}. The interface shows the lightcurve folded at the best period (top left) with individual data points in black and a binning in yellow, the BLS spectrum (top right), the full lightcurve (middle left) where the transits are indicated in green, a zoom on the transit (middle right), the centroid variations, a visible and an infrared image of the star's neighborhood, the results of a transit fit, and a list of parameters (BLS outputs, transit parameters, magnitudes, color, spectral type, Simbad information, etc.).}
\label{fig: xo poster}
\end{figure}

\begin{figure}
\includegraphics[width=5.7cm]{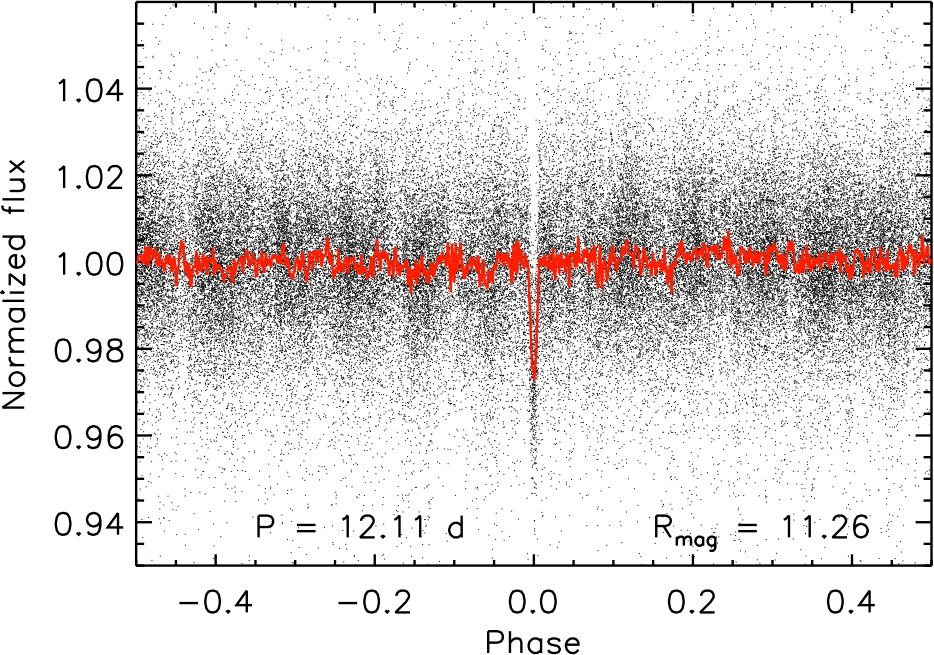}\hfill
\includegraphics[width=5.7cm]{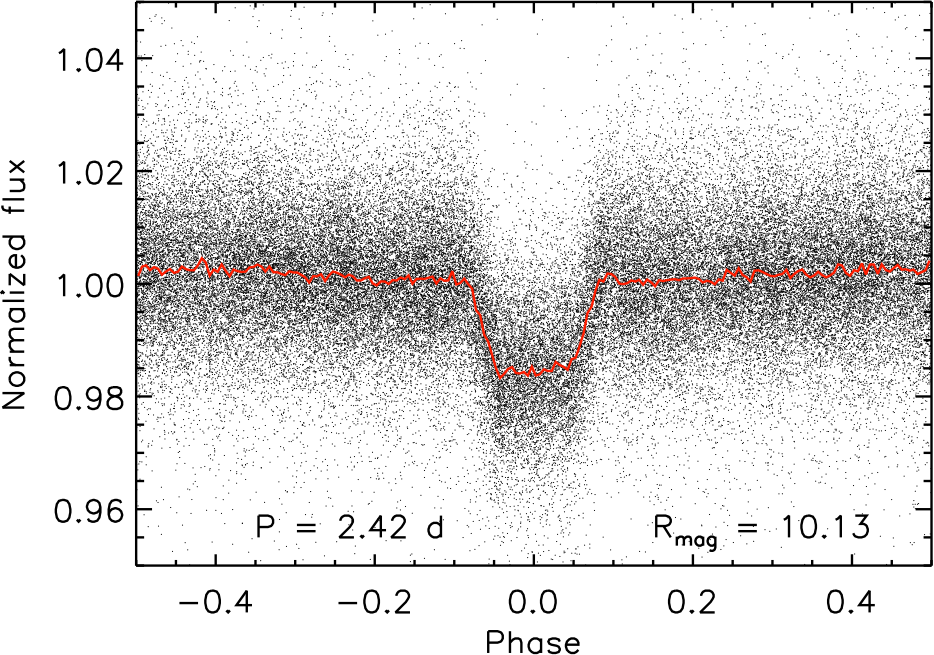}
\includegraphics[width=5.7cm]{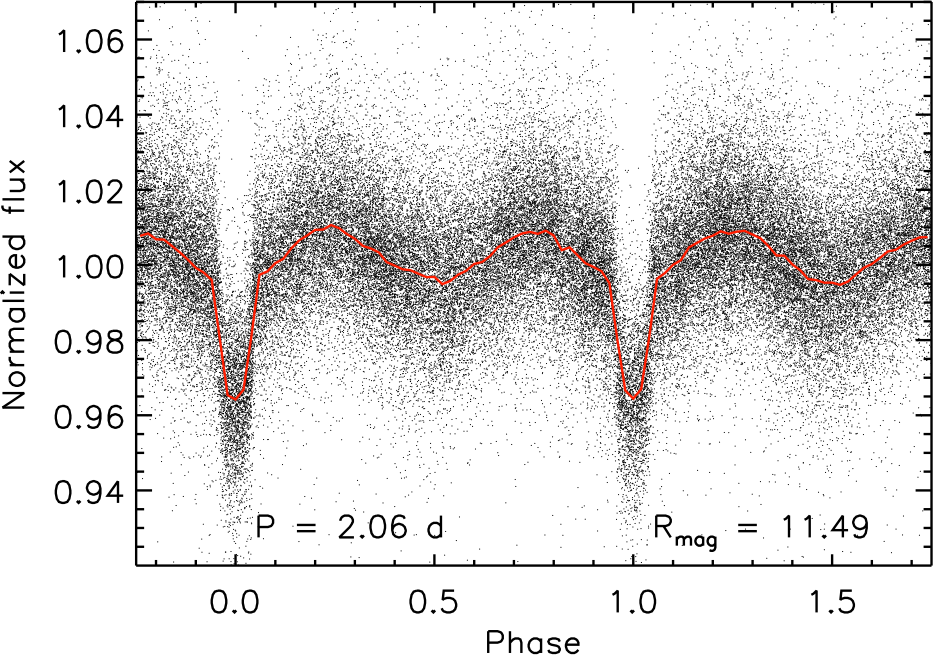}\hfill
\includegraphics[width=5.7cm]{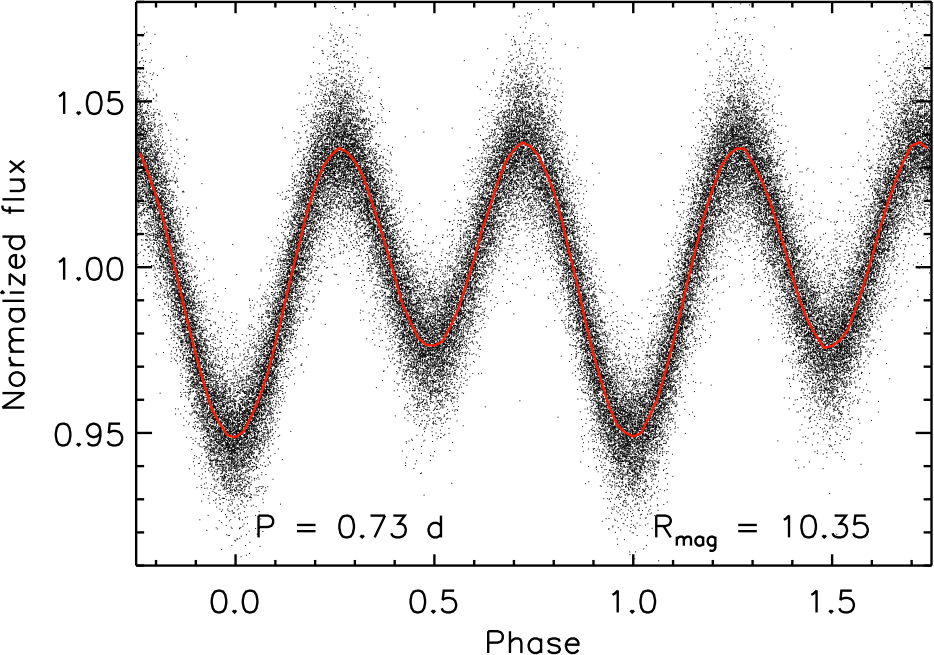}
\includegraphics[width=5.7cm]{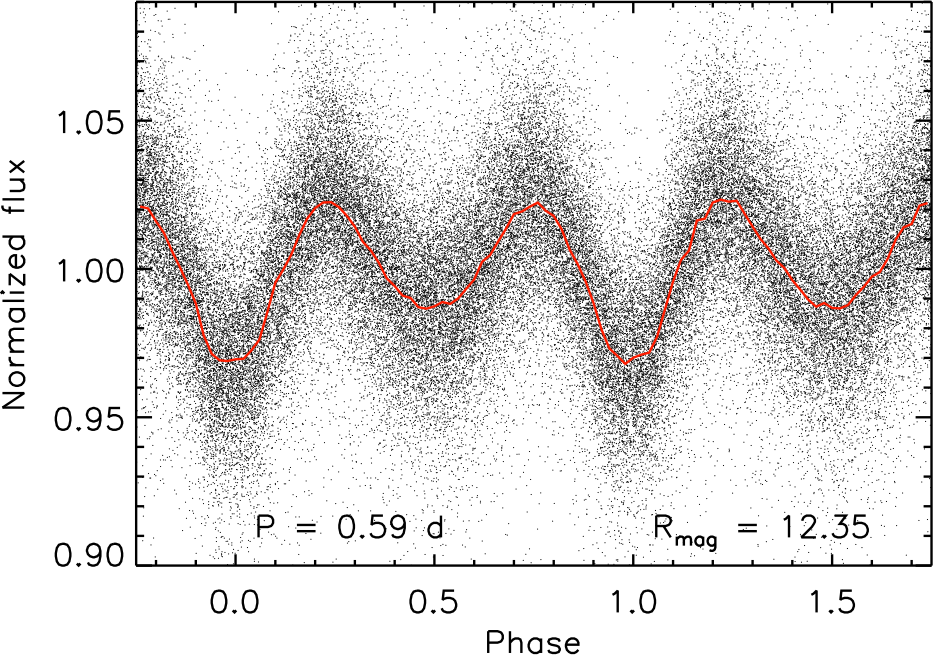}\hfill
\includegraphics[width=5.7cm]{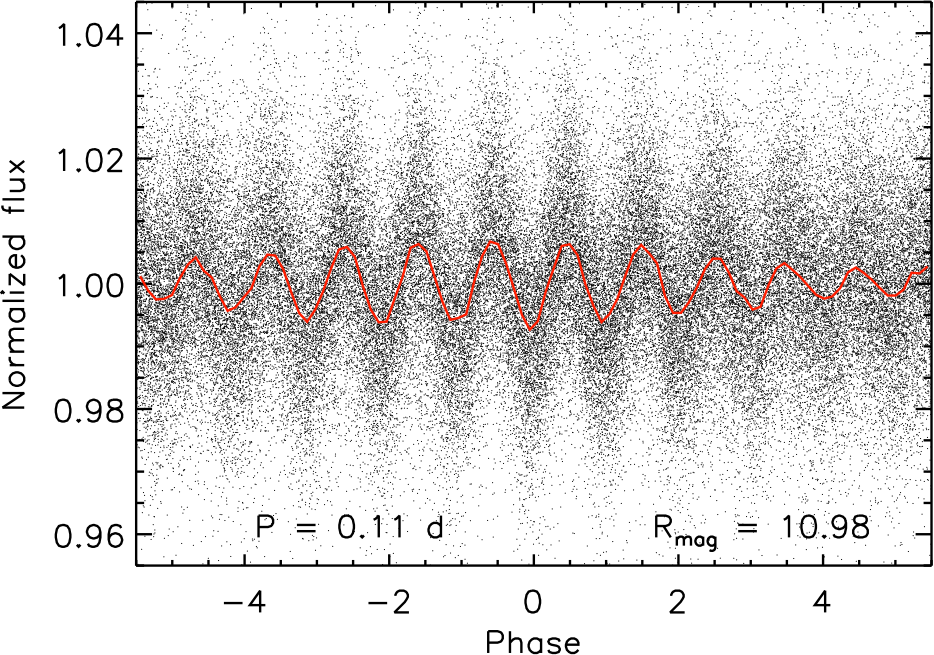}
\includegraphics[width=5.7cm]{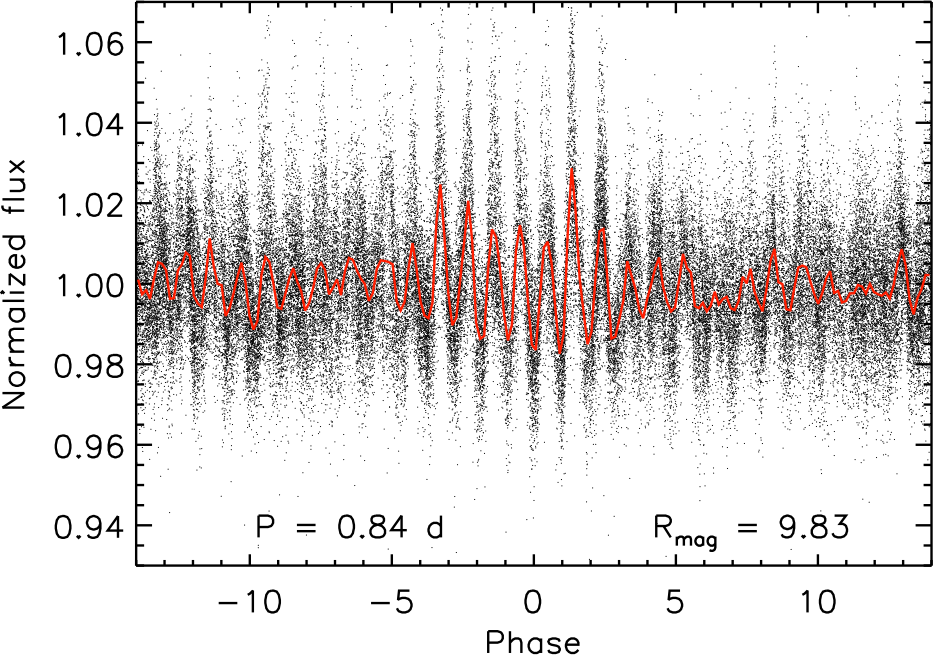}\hfill       
\includegraphics[width=5.7cm]{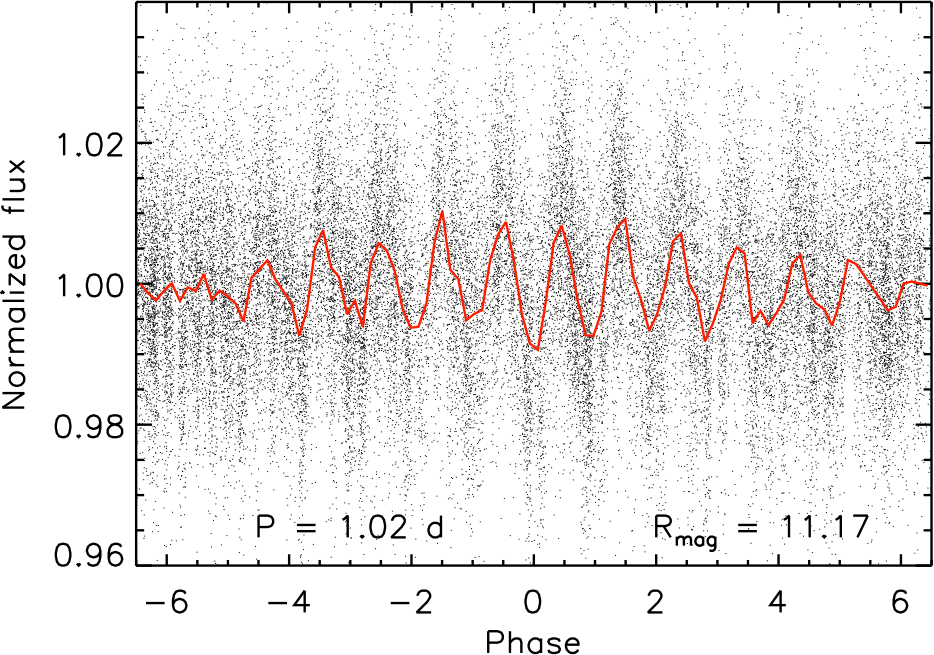}
\caption{Lightcurves of eclipsing objects and variable stars with short periods obtained with XO. Individual data points obtained with a time interval of 6.3 minutes are shown in black and a binning is shown in red. The period in days and the $R$ magnitude are indicated on each plot. From left to right and top to bottom: a transiting planet candidate with a 12 day period, a probable eclipsing binary, a semi-detached eclipsing binary, two variable stars, three variable stars with beats (with the lowest period indicated).}
\label{fig: lightcurves}
\end{figure}

\begin{figure}
\includegraphics[width=5.7cm]{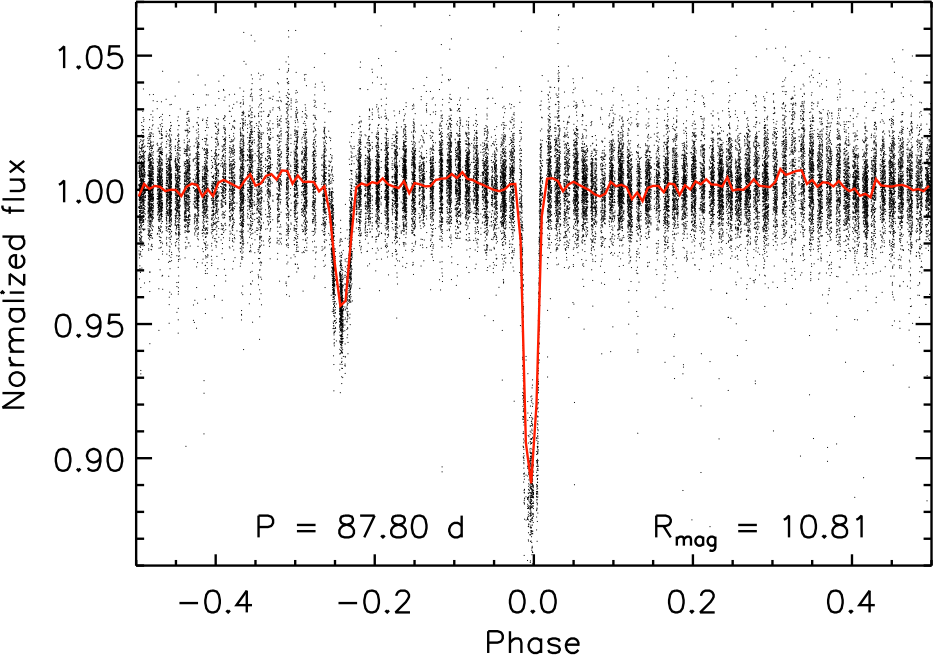}\hfill
\includegraphics[width=5.7cm]{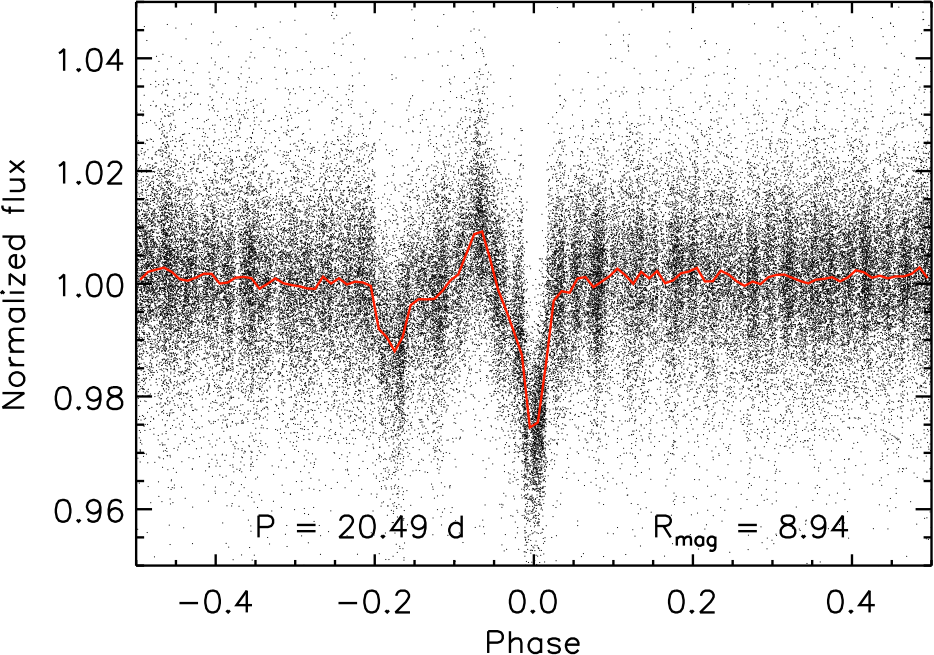}
\includegraphics[width=5.7cm]{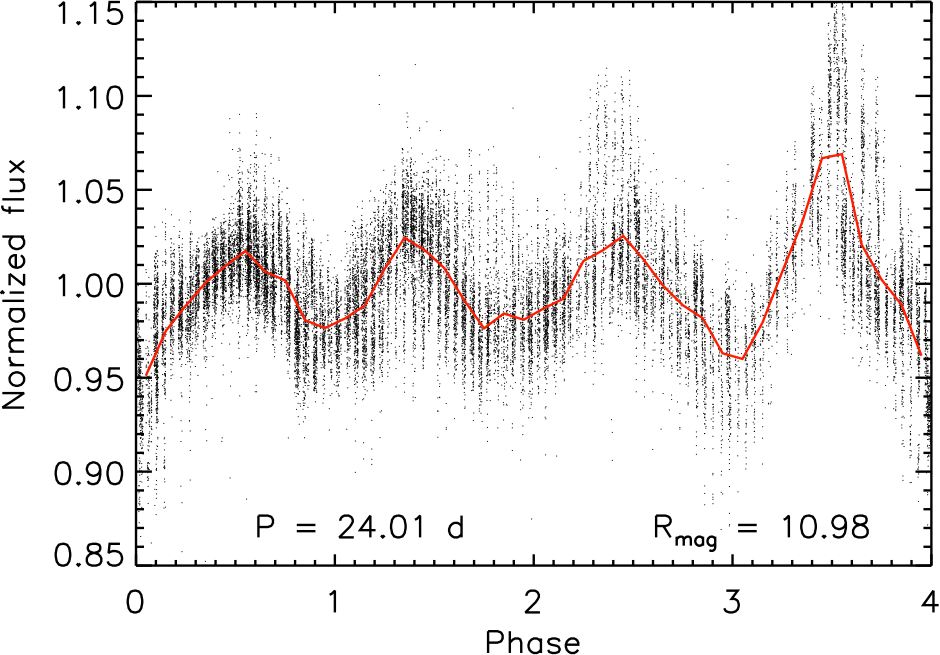}\hfill
\includegraphics[width=5.7cm]{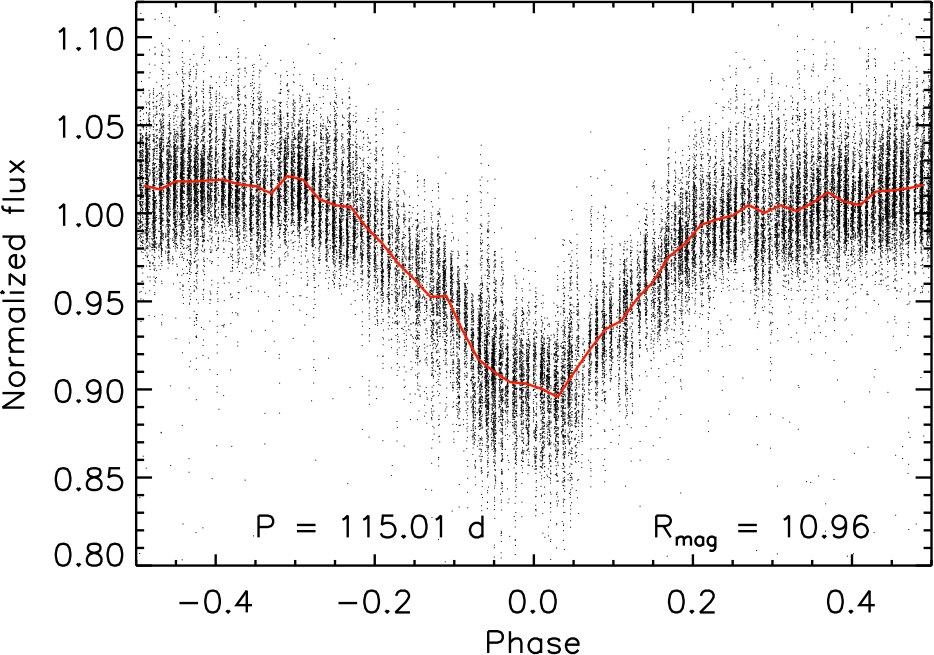}
\includegraphics[width=5.7cm]{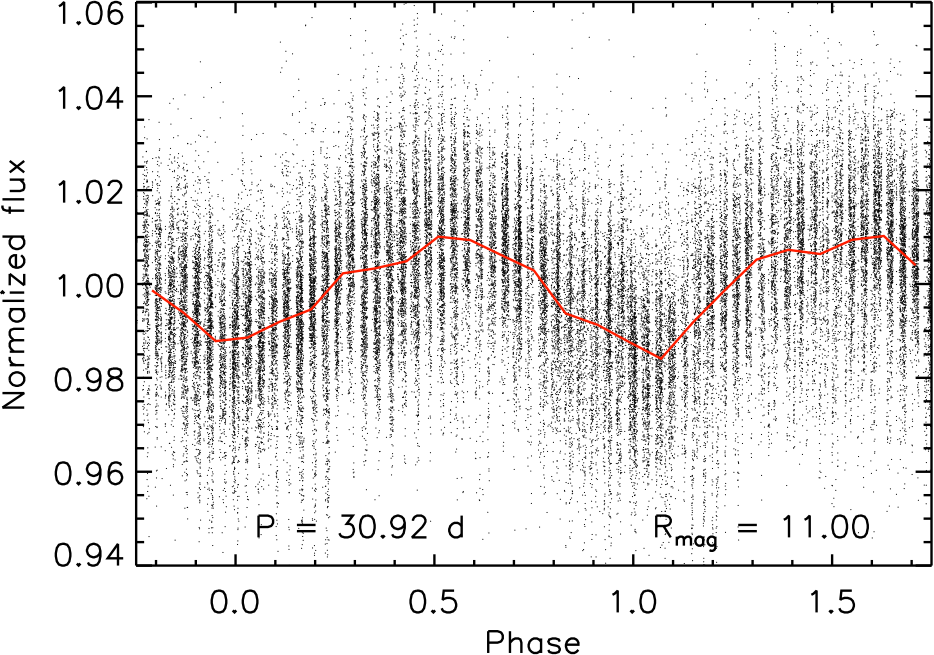}\hfill
\includegraphics[width=5.7cm]{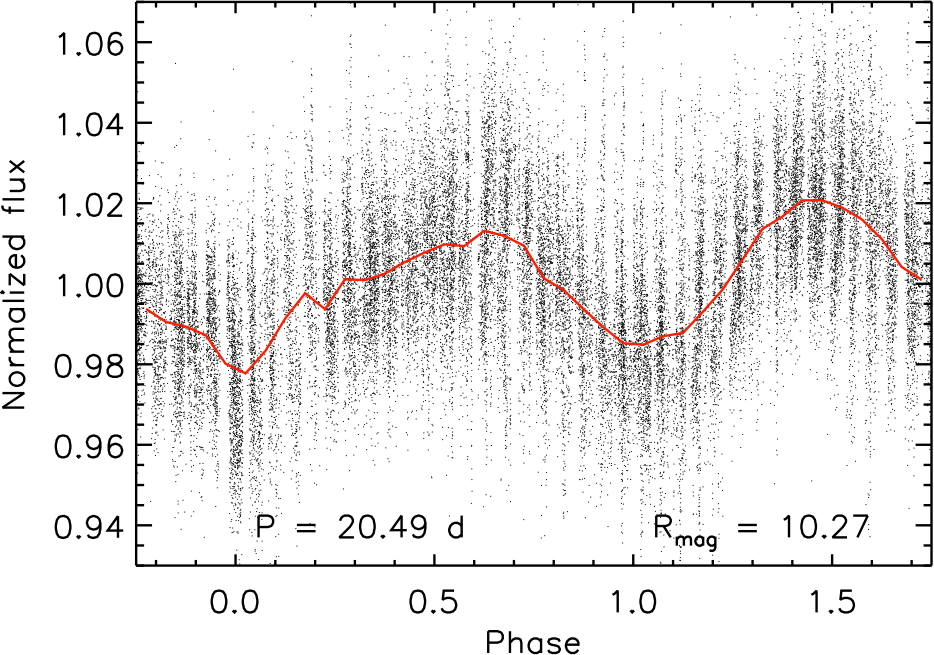}       
\caption{Same as Figure \ref{fig: lightcurves} for long period objects. From left to right and top to bottom: an eccentric eclipsing binary with a 88 day period, a heartbeat eccentric eclipsing binary with a 20 day period, a 24 day period variable star with amplitude variations, and three variable stars with 115 day, 31 day, and 20 day periods.}
\label{fig: lightcurves long}
\end{figure}

Finally, we combine the 15 intervals into 3 period ranges: 0.4 -- 1 day, 1 -- 10 days, and 10 -- 100 days. For each lightcurve, we compare the results obtained for the intervals within each range and keep the period that yields the largest power in the BLS spectrum.
All the lightcurves end up with a best period but most of them do not contain a valid signal. We keep only those with $\alpha > 15$ and sort them by decreasing $\alpha$ (lightcurves with the most significant signals are ranked first). This yields around 300 lightcurves out of 2000 for each field. We developed an interactive program in $IDL$ to inspect these selected lightcurves and to identify transit candidates, as displayed in Figure \ref{fig: xo poster}. From this search, we identified hundreds of variable stars and a few tens of transiting planet candidates. Examples of lightcurves of variable objects are shown in Figures~\ref{fig: lightcurves} and ~\ref{fig: lightcurves long}.

To evaluate the transit search efficiency, we inject fake transits in the lightcurves. We choose 20 lightcurves of stars of various magnitudes and assign them different periods as well as transit parameters randomly chosen within physically plausible ranges. We also inject the transits of the planets discovered by the WASP and HAT surveys in the lightcurves of stars of similar magnitudes (these planets have periods generally below 10 days). This yields a total of 167 lightcurves with injected transits that we concatenate at the end of the star -- epoch magnitude array, and that we analyze in the same way as regular lightcurves. More than 90\% of the injected planets are recovered after the BLS search and the transit candidate selection.

\subsection{Follow-up observations}
\label{sec: Follow-up observations}

Follow-up observations are necessary to confirm or reject transit candidates. We built a large follow-up team of amateur and professional astronomers who run these observations using facilities reported in Table \ref{tab: follow-up facilities}. We perform photometry at the predicted transit ephemerides to check the reality of the signals; most candidates are not detected and are rejected. Then, we perform multi-color photometry: we observe transits in different bandpasses to identify eclipsing binaries, which have a color-dependent transit depth. We also compare the odd and even transit depths, which usually differ for eclipsing binaries, and search for secondary eclipses that may be unseen in the XO lightcurves. Then, valid candidates are sent for radial velocity observations with the SOPHIE spectrograph at the Observatoire de Haute-Provence, France \citep{Bouchy2009}. In practice, follow-up observations are very time consuming, so the candidates are ordered by priority and the type of follow-up observation is carefully chosen. For example, a few candidates show a very clear signal in the XO data and have all the properties of a transiting planet; these are sent directly for radial velocities. Others are sent to multi-color photometry: one transit observed in two colors can be enough to identify an eclipsing binary. Finally, some objects require specific observations. This was the case for the hot Jupiter XO-6b which orbits a fast rotating F5 star: the radial velocity precision was limited to about 70 $\rm m \,s^{-1}$ due to the stellar rotation, which was too large to confirm the presence of the planet. We performed Rossiter-McLaughlin observations with SOPHIE and detected the planet's signature in Doppler tomography, which confirmed its planetary nature \citep{Crouzet2017}. Follow-up observations of other transit candidates are underway.

\begin{table}
\begin{center}
\caption{Observatories and telescopes used for follow-up observations of XO transit candidates. The diameters of the primary mirrors are given in inches and cm.}
\label{tab: follow-up facilities}
\begin{tabular}{lll}
\hline
\hline
Observatory & Telescope & Purpose  \\
\hline
Observatoire de Haute-Provence, France & 76" (193 cm), SOPHIE spectrograph  &  Radial velocities \\
Hereford Arizona Observatory, Arizona, USA & Celestron 11" (28 cm), Meade 14" (36 cm)  & Photometry \\
Acton Sky Portal, Massachusetts, USA & 11" (28 cm)  &  Photometry  \\
Observatori Astron\`omic del Montsec, Catalonia, Spain & Joan Or\'o Telescope 31" (80 cm)  & Photometry  \\
Observatoire de Nice, France & Schaumasse 16" (40 cm)  &  Photometry  \\
Vermillion Cliffs Observatory, Kanab, Utah, USA & 24'' (60 cm)  &  Photometry \\
Elgin Observatory, Elgin, Oregon, USA & 12" (30 cm)  &  Photometry \\
\hline
\hline
\end{tabular}
\end{center}
\end{table}

\subsection{Planet detection yield and discoveries}
\label{sec: Yield and planet detection}

We estimate the yield of exoplanet discovery with XO and compare it to the number of detected planets. The aim is to inform us on the observation and data analysis efficiency and not to perform an accurate completeness study. Thus, we compute only simple estimates.
We use the yield simulations developed by \citet{Sullivan2015} for the NASA \tess mission \citep[\textit{Transiting Exoplanet Survey Satellite,}][]{Ricker2014}, which in turn are based off results from the \kepler mission. We consider only planets with orbital periods $P < 20$ days, because they are easier to detect due to the decreasing transit probability for longer orbital periods, as evidenced by the WASP \& HAT planet distributions peaking around 3-4 days.
The \tess simulations cover 95\% of the entire sky, \ie 38,000\degreesq. The number of transiting planets with $P < 20$ days and radii greater than four times that of Earth ($R_p > 0.4 \;R_{Jup}$) orbiting stars with Cousins $I$ band magnitude $I_c < 10$ over this area is expected to be \simi100, and \simi300 for $I_c < 11$ \citep[][Figures 11 \& 22]{Sullivan2015}. The XO units concentrate on 520\degreesq (10 fields of 52\degreesq each). Thus, the yield of such planets is 1.4 for $I_c < 10$ and 4.1 for $I_c < 11$.
In this calculation we assume ideal monitoring, \ie no limitation due to the observing window or instrumental precision. For \tess, this is well justified as the monitoring will be continuous for at least 30 days with a precision that is much better than necessary for the detection of transiting giant exoplanets (in other words, \tess will detect nearly every planet with those characteristics, at least around the selected target stars). For XO, the observing window per field extends to approximately the nighttime of eighteen months (twice nine months, two strips, two longitudes considering that Observatorio del Teide and Observatori Astron\`omic del Montsec have a similar longitude), and the precision of the full lightcurves folded at short periods and binned on timescales that are relevant to transits is at the millimagnitude level (see Section \textit{\nameref{sec: Photometric precision}} and Figures \ref{fig: xo poster} and \ref{fig: lightcurves}). Thus, transiting close-in giant planets should be detected, if present. 
We found two such planets with the XO units so far, XO-6b \citep{Crouzet2017} and XO-7b \citep{Crouzet-inprep}, which host stars have $I_c$ magnitudes of 10.12 and 10.27 respectively. This number is in line with the approximate yield. Thus, the observations and data analysis are as efficient as one could expect. A re-analysis of the XO data to find new transiting close-in gas giant planets would be of low gain and we will simply pursue the follow-up observations of the candidates we have identified. Although transiting planets with longer periods are more challenging to detect, we found a few candidates with $P > 20$ days which are also under follow-up observations.

\section{Conclusion}
\label{sec: Conclusion}

In this chapter, we presented an overview of the second version of the XO project: instrumental setup, operations, data reduction, instrumental performances, search for transit signals, planet yield and discoveries. We observed two strips covering an effective sky area of $520^{\circ2}$ for twice nine months using the CCDs in time-delayed integration, and we extracted the lightcurves of \simi20,000 bright stars up to magnitude $R \approx 12.5$. The precision is at the millimagnitude level when folding the lightcurves on timescales that are relevant to transits. In addition, this setup allows us to detect long period signals, up to $P \approx 100$ days. We identified several hundreds of variable stars and a few tens of planet candidates. The transiting hot Jupiter \mbox{XO-6b} orbiting a fast rotating star has been discovered from this work \citep{Crouzet2017}. Another planet, XO-7b, has been confirmed \citep{Crouzet-inprep} and other transit candidates are under follow-up observations. The XO observations have been discontinued in anticipation of the NASA \tess mission and the work on XO is now dedicated to the follow-up and study of individual objects.

\section{Acknowledgments}
N.C. gratefully acknowledges Peter R. McCullough as the founder and Principal Investigator of the XO project. The XO project was supported by NASA grant NNX10AG30G.
This research has made use of the Extrasolar Planets Encyclopaedia (exoplanet.eu) and the Simbad database (simbad.u-strasbg.fr/simbad/).
Software: astrometry.net \citep{Lang2010}, Stellar Photometry Software \citep{Janes1993}.

\bibliographystyle{spbasicHBexo}

\begin{thebibliography}{20}
\providecommand{\natexlab}[1]{#1}
\providecommand{\url}[1]{{#1}}
\providecommand{\urlprefix}{URL }
\expandafter\ifx\csname urlstyle\endcsname\relax
  \providecommand{\doi}[1]{DOI~\discretionary{}{}{}#1}\else
  \providecommand{\doi}{DOI~\discretionary{}{}{}\begingroup
  \urlstyle{rm}\Url}\fi
\providecommand{\eprint}[2][]{\url{#2}}

\bibitem[{{Bakos} et~al.(2004){Bakos}, {Noyes}, {Kov{\'a}cs}, {Stanek},
  {Sasselov}, and {Domsa}}]{Bakos2004}
{Bakos} G, {Noyes} RW, {Kov{\'a}cs} G et~al. (2004) {Wide-Field Millimagnitude
  Photometry with the HAT: A Tool for Extrasolar Planet Detection}. \pasp
  116:266--277

\bibitem[{{Bakos} et~al.(2002){Bakos}, {L{\'a}z{\'a}r}, {Papp}, {S{\'a}ri}, and
  {Green}}]{Bakos2002}
{Bakos} G{\'A}, {L{\'a}z{\'a}r} J, {Papp} I, {S{\'a}ri} P {Green} EM (2002)
  {System Description and First Light Curves of the Hungarian Automated
  Telescope, an Autonomous Observatory for Variability Search}. \pasp
  114:974--987

\bibitem[{{Bouchy} et~al.(2009){Bouchy}, {H{\'e}brard}, {Udry}, {Delfosse},
  {Boisse}, {Desort}, {Bonfils}, {Eggenberger}, {Ehrenreich}, {Forveille},
  {Lagrange}, {Le Coroller}, {Lovis}, {Moutou}, {Pepe}, {Perrier}, {Pont},
  {Queloz}, {Santos}, {S{\'e}gransan}, and {Vidal-Madjar}}]{Bouchy2009}
{Bouchy} F, {H{\'e}brard} G, {Udry} S et~al. (2009) {The SOPHIE search for
  northern extrasolar planets . I. A companion around HD 16760 with mass close
  to the planet/brown-dwarf transition}. \aap 505:853--858

\bibitem[{{Burke} et~al.(2007){Burke}, {McCullough}, {Valenti}, {Johns-Krull},
  {Janes}, {Heasley}, {Summers}, {Stys}, {Bissinger}, {Fleenor}, {Foote},
  {Garc{\'{\i}}a-Melendo}, {Gary}, {Howell}, {Mallia}, {Masi}, {Taylor}, and
  {Vanmunster}}]{Burke2007}
{Burke} CJ, {McCullough} PR, {Valenti} JA et~al. (2007) {XO-2b: Transiting Hot
  Jupiter in a Metal-rich Common Proper Motion Binary}. \apj 671:2115--2128

\bibitem[{{Burke} et~al.(2008){Burke}, {McCullough}, {Valenti}, {Long},
  {Johns-Krull}, {Machalek}, {Janes}, {Taylor}, {Fleenor}, {Foote}, {Gary},
  {Garc{\'{\i}}a-Melendo}, {Gregorio}, and {Vanmunster}}]{Burke2008}
{Burke} CJ, {McCullough} PR, {Valenti} JA et~al. (2008) {XO-5b: A Transiting
  Jupiter-sized Planet with a 4 day Period}. \apj 686:1331--1340

\bibitem[{{Collier Cameron} et~al.(2007){Collier Cameron}, {Bouchy},
  {H{\'e}brard}, {Maxted}, {Pollacco}, {Pont}, {Skillen}, {Smalley}, {Street},
  {West}, {Wilson}, {Aigrain}, {Christian}, {Clarkson}, {Enoch}, {Evans},
  {Fitzsimmons}, {Fleenor}, {Gillon}, {Haswell}, {Hebb}, {Hellier}, {Hodgkin},
  {Horne}, {Irwin}, {Kane}, {Keenan}, {Loeillet}, {Lister}, {Mayor}, {Moutou},
  {Norton}, {Osborne}, {Parley}, {Queloz}, {Ryans}, {Triaud}, {Udry}, and
  {Wheatley}}]{Collier2007}
{Collier Cameron} A, {Bouchy} F, {H{\'e}brard} G et~al. (2007) {WASP-1b and
  WASP-2b: two new transiting exoplanets detected with SuperWASP and SOPHIE}.
  \mnras 375:951--957

\bibitem[{{Crouzet} et~al.(2017){Crouzet}, {McCullough}, {Long}, {Montanes
  Rodriguez}, {Lecavelier des Etangs}, {Ribas}, {Bourrier}, {H{\'e}brard},
  {Vilardell}, {Deleuil}, {Herrero}, {Garcia-Melendo}, {Akhenak}, {Foote},
  {Gary}, {Benni}, {Guillot}, {Conjat}, {M{\'e}karnia}, {Garlitz}, {Burke},
  {Courcol}, and {Demangeon}}]{Crouzet2017}
{Crouzet} N, {McCullough} PR, {Long} D et~al. (2017) {Discovery of XO-6b: A Hot
  Jupiter Transiting a Fast Rotating F5 Star on an Oblique Orbit}. \aj 153:94

\bibitem[{{Crouzet} et~al.(in prep){Crouzet}, {McCullough}, {Long}, {Montanes
  Rodriguez}, {Lecavelier des Etangs}, {Ribas}, {Bourrier}, {H{\'e}brard},
  {Vilardell}, {Deleuil}, {Herrero}, {Garcia-Melendo}, {Akhenak}, {Foote},
  {Gary}, {Benni}, {Guillot}, {Conjat}, {M{\'e}karnia}, {Garlitz}, {Burke},
  {Courcol}, and {Demangeon}}]{Crouzet-inprep}
{Crouzet} N, {McCullough} PR, {Long} D et~al. (in prep) {Discovery of XO-7b}

\bibitem[{{Janes} and {Heasley}(1993)}]{Janes1993}
{Janes} KA {Heasley} JN (1993) {Stellar Photometry Software}. \pasp
  105:527--537

\bibitem[{{Johns-Krull} et~al.(2008){Johns-Krull}, {McCullough}, {Burke},
  {Valenti}, {Janes}, {Heasley}, {Prato}, {Bissinger}, {Fleenor}, {Foote},
  {Garcia-Melendo}, {Gary}, {Howell}, {Mallia}, {Masi}, and
  {Vanmunster}}]{JohnsKrull2008}
{Johns-Krull} CM, {McCullough} PR, {Burke} CJ et~al. (2008) {XO-3b: A Massive
  Planet in an Eccentric Orbit Transiting an F5 V Star}. \apj 677:657--670

\bibitem[{{Kov{\'a}cs} et~al.(2002){Kov{\'a}cs}, {Zucker}, and
  {Mazeh}}]{Kovacs2002}
{Kov{\'a}cs} G, {Zucker} S {Mazeh} T (2002) {A box-fitting algorithm in the
  search for periodic transits}. \aap 391:369--377

\bibitem[{{Lang} et~al.(2010){Lang}, {Hogg}, {Mierle}, {Blanton}, and
  {Roweis}}]{Lang2010}
{Lang} D, {Hogg} DW, {Mierle} K, {Blanton} M {Roweis} S (2010) {Astrometry.net:
  Blind Astrometric Calibration of Arbitrary Astronomical Images}. \aj
  139:1782--1800

\bibitem[{{McCullough} et~al.(2005){McCullough}, {Stys}, {Valenti}, {Fleming},
  {Janes}, and {Heasley}}]{McCullough2005}
{McCullough} PR, {Stys} JE, {Valenti} JA et~al. (2005) {The XO Project:
  Searching for Transiting Extrasolar Planet Candidates}. \pasp 117:783--795

\bibitem[{{McCullough} et~al.(2006){McCullough}, {Stys}, {Valenti},
  {Johns-Krull}, {Janes}, {Heasley}, {Bye}, {Dodd}, {Fleming}, {Pinnick},
  {Bissinger}, {Gary}, {Howell}, and {Vanmunster}}]{McCullough2006}
{McCullough} PR, {Stys} JE, {Valenti} JA et~al. (2006) {A Transiting Planet of
  a Sun-like Star}. \apj 648:1228--1238

\bibitem[{{McCullough} et~al.(2008){McCullough}, {Burke}, {Valenti}, {Long},
  {Johns-Krull}, {Machalek}, {Janes}, {Taylor}, {Gregorio}, {Foote}, {Gary},
  {Fleenor}, {Garc{\'{\i}}a-Melendo}, and {Vanmunster}}]{McCullough2008}
{McCullough} PR, {Burke} CJ, {Valenti} JA et~al. (2008) {XO-4b: An Extrasolar
  Planet Transiting an F5V Star}. ArXiv e-prints

\bibitem[{{Pollacco} et~al.(2006){Pollacco}, {Skillen}, {Collier Cameron},
  {Christian}, {Hellier}, {Irwin}, {Lister}, {Street}, {West}, {Anderson},
  {Clarkson}, {Deeg}, {Enoch}, {Evans}, {Fitzsimmons}, {Haswell}, {Hodgkin},
  {Horne}, {Kane}, {Keenan}, {Maxted}, {Norton}, {Osborne}, {Parley}, {Ryans},
  {Smalley}, {Wheatley}, and {Wilson}}]{Pollacco2006}
{Pollacco} DL, {Skillen} I, {Collier Cameron} A et~al. (2006) {The WASP Project
  and the SuperWASP Cameras}. \pasp 118:1407--1418

\bibitem[{{Ricker} et~al.(2014){Ricker}, {Winn}, {Vanderspek}, {Latham},
  {Bakos}, {Bean}, {Berta-Thompson}, {Brown}, {Buchhave}, {Butler}, {Butler},
  {Chaplin}, {Charbonneau}, {Christensen-Dalsgaard}, {Clampin}, {Deming},
  {Doty}, {De Lee}, {Dressing}, {Dunham}, {Endl}, {Fressin}, {Ge}, {Henning},
  {Holman}, {Howard}, {Ida}, {Jenkins}, {Jernigan}, {Johnson}, {Kaltenegger},
  {Kawai}, {Kjeldsen}, {Laughlin}, {Levine}, {Lin}, {Lissauer}, {MacQueen},
  {Marcy}, {McCullough}, {Morton}, {Narita}, {Paegert}, {Palle}, {Pepe},
  {Pepper}, {Quirrenbach}, {Rinehart}, {Sasselov}, {Sato}, {Seager},
  {Sozzetti}, {Stassun}, {Sullivan}, {Szentgyorgyi}, {Torres}, {Udry}, and
  {Villasenor}}]{Ricker2014}
{Ricker} GR, {Winn} JN, {Vanderspek} R et~al. (2014) {Transiting Exoplanet
  Survey Satellite (TESS)}. In: Space Telescopes and Instrumentation 2014:
  Optical, Infrared, and Millimeter Wave, \procspie, vol 9143, p 914320,
  \doi{10.1117/12.2063489}

\bibitem[{{Santerne} et~al.(2016){Santerne}, {Moutou}, {Tsantaki}, {Bouchy},
  {H{\'e}brard}, {Adibekyan}, {Almenara}, {Amard}, {Barros}, {Boisse},
  {Bonomo}, {Bruno}, {Courcol}, {Deleuil}, {Demangeon}, {D{\'{\i}}az},
  {Guillot}, {Havel}, {Montagnier}, {Rajpurohit}, {Rey}, and
  {Santos}}]{Santerne2016}
{Santerne} A, {Moutou} C, {Tsantaki} M et~al. (2016) {SOPHIE velocimetry of
  Kepler transit candidates. XVII. The physical properties of giant exoplanets
  within 400 days of period}. \aap 587:A64

\bibitem[{{Sullivan} et~al.(2015){Sullivan}, {Winn}, {Berta-Thompson},
  {Charbonneau}, {Deming}, {Dressing}, {Latham}, {Levine}, {McCullough},
  {Morton}, {Ricker}, {Vanderspek}, and {Woods}}]{Sullivan2015}
{Sullivan} PW, {Winn} JN, {Berta-Thompson} ZK et~al. (2015) {The Transiting
  Exoplanet Survey Satellite: Simulations of Planet Detections and
  Astrophysical False Positives}. \apj 809:77

\bibitem[{{Tamuz} et~al.(2005){Tamuz}, {Mazeh}, and {Zucker}}]{Tamuz2005}
{Tamuz} O, {Mazeh} T {Zucker} S (2005) {Correcting systematic effects in a
  large set of photometric light curves}. \mnras 356:1466--1470

\end{thebibliography}

\end{document}